# Frequency Dispersion of Sound Propagation in Rouse Polymer Melts via Generalized Dynamic Random Phase Approximation

Igor Ya. Erukhimovich[1,*] and Yaroslav V. Kudryavtsev[2]

[1] Physics Department, Moscow State University,
Leninskie Gory, Moscow 119992, Russia;
ierukhs@polly.phys.msu.ru;

[2] Topchiev Institute of Petrochemical Synthesis,
Russian Academy of Sciences,
Leninsky prospect 29, Moscow 119991, Russia;
yar@ips.ac.ru;

**Shortened title:**

I. Erukhimovich et al. Dispersion of Sound in Polymer Melts via DRPA

**Abstract.** An extended generalization of the dynamic random phase approximation (DRPA) for $L$-component polymer systems is presented. Unlike the original version of the DRPA, which relates the $(L \times L)$ matrices of the collective density-density time correlation functions and the corresponding susceptibilities of concentrated polymer systems to those of the tracer macromolecules and so-called broken links system (BLS), our generalized DRPA solves this problem for the $(5 \times L) \times (5 \times L)$ matrices of the coupled susceptibilities and time correlation functions of the component number, kinetic energy and flux densities. The presented technique is used to study propagation of sound and dynamic form-factor in disentangled (Rouse) monodisperse homopolymer melt. The calculated ultrasonic velocity and absorption coefficient reveal substantial frequency dispersion. The relaxation time $\tau$ is proportional to the degree of polymerization $N$, which is $N$ times less than the Rouse time and evidences strong dynamic screening because of interchain interaction. We discuss also some peculiarities of the Brillouin scattering in polymer melts. Besides, a new convenient expression for the dynamic structure function of the single Rouse chain in $(\mathbf{q}, p)$− representation is found.

---

* To whom correspondence should be addressed

# 1 Introduction

The purpose of the present paper is two-fold. In terms of the final output, we deal here with the sound (ultrasonic) propagation via the monodisperse melt of homopolymer linear chains and calculate the dependence of both the acoustic attenuation coefficient α and sound velocity $v$ on the sound frequency $f$. Such a calculation is interesting by itself and important to interpret the results of ultrasonic measurements in a due way.

Indeed, a substantial frequency dispersion of the sound absorption coefficient $\alpha/f^2$ in polymer melts was predicted theoretically [1-4] and found experimentally in the ultrasonic range [5]. Gotlib and Salikhov [1] modeled the behavior of a concentrated solution by a three-dimensional network of polymer subchains immersed in a solvent. They found that the absorption coefficient $\alpha/f^2$ scales as $f^{-1/2}$ at $f \sim 10^6$-$10^7$ Hz, which is consistent with the frequency dependence of the shear viscosity in the Rouse model [6]. A somewhat stronger decrease of $\alpha/f^2$ observed experimentally at higher frequencies (up to $10^9$ Hz) has been attributed by Gotlib and Darinskii [2] to the finiteness of the relaxation spectrum for Gaussian subchains between the network knots. Berger and Straube [3] have calculated the sound absorption coefficient by solving the diffusion equation for the distribution function of the structural units (segments) of a single polymer chain interacting with a solvent (dilute solution) and found, in the megahertz range, the Rouse-type behavior $\alpha/f^2 \sim f^{-1/2}$. They also concluded [4] that the temporary entanglements present between the chains in concentrated solutions change the dependence of $\alpha/f^2$ to $f^{-3/4}$.

However, all the mentioned theoretical studies neglect the inter-chain interactions and, thus, are in principle not capable of studying the frequency dispersion of the sound velocity $v$ directly related to the system compressibility. Instead, they assume the sound absorption to occur independently on the separate polymer chains due to the viscous interactions between polymer chains immersed in a low-molecular solvent. Meanwhile, no substantial difference was found between the solvent and melt-type behavior of acoustic attenuation [5,7]. Thus, it is reasonable to suppose that the direct interactions between polymer chains in absence of any solvent also would lead to a dispersion of $\alpha/f^2$. Besides, it is natural to expect the value of $v$ for polymer systems to be determined by interplay of all the relaxation processes (mass, flux and heat transfer) since even for simple liquids $v$ depends on the adiabatic (rather than isothermic) compressibility [8].

Many authors addressed the problem of self-consistent simultaneous description of all the aforementioned relaxation processes in polymers. Often one takes into account the existence of a non-zero velocity field in polymer solutions via introducing hydrodynamic interactions between polymer chains described by the configuration-dependent Oseen tensor [6]. Along this line, the influence of the hydrodynamic interactions on self-diffusion [9], spinodal decomposition [10,11] and diffusive relaxation in multicomponent systems [12] was investigated. Spinodal decomposition with respect to the free volume non-homogeneous distribution was considered using simultaneous equations of the polymer density and flux relaxation [13]. The stress imposed by the network of entanglements in a concentrated solution of long chains also affects the density relaxation [14]. The influence of viscoelasticity on the self- and mutual diffusion was studied in Ref. [15]. But in all cited papers the coupling between the concentration and velocity relaxation was considered in the isothermic limit, i.e., the processes of thermal relaxation were ignored, which, as mentioned above, is definitely incorrect when considering the sound velocity dispersion.

Another important shortcoming of the aforementioned approaches is that neither of them guarantees that the correlation functions and the dissipative properties of the polymer systems under study do obey the famous fluctuation-dissipation theorem [16,17]. The only approach free of this shortcoming just by its construction is the dynamic random phase approximation (DRPA) [18-20] we believe to be the most elaborated and consequent way to study the dynamical properties of the concentrated polymer solutions and melts (in general, polydisperse). In particular, it is within the DRPA that the "breathing" mode in block copolymers [19] and large-scale "composition" relaxation in polydisperse blends of homopolymers [20] were predicted and described as consistent with the experimental data (see Refs. [21-27] and references therein). (Some authors [28-30] claimed that the DRPA couldn't explain the "fast mode diffusion behavior", which is sometimes observed experimentally. However, the more thorough analysis has shown that such a behavior could be perfectly described within the DRPA when taking into account the presence of vacancies [31,32] or the finite compressibility effects [33].)

But the original version of the DRPA [18-20] was also elaborated to deal with the concentration relaxation in polymers in the isothermic-isobaric limit. Even though Akcasu et. al. have included into the DRPA the hydrodynamic interactions between polymer chains via the Oseen tensor [12], this *ad hoc* ansatz needs an additional validation.

Thus, another (and, perhaps, even more fundamental) aim of our paper is to present the due generalization of the DRPA enabling us to properly take into account the coupling between the density, temperature and flux fluctuations. For this purpose, the paper is organized as follows. In section 2 we remind the reader the basic concepts of the generalized hydrodynamics, which is closely related to the DRPA. In section 3 we first develop the dynamic RPA beyond the framework of pure diffusion and introduce the central notions of the structural and direct susceptibilities. In section 4 we calculate the structural susceptibility for a multicomponent polymer blend. The direct susceptibility is calculated in section 5. The detailed analysis of the generalized susceptibility with regard to the sound propagation in a one-component compressible polymer melt is presented in section 6. In the last section conclusions are formulated. The Appendix contains a new method of calculating the dynamic correlators for the Rouse model.

## 2  Generalized hydrodynamics

A conventional hydrodynamic description for any system is based on the assumption of a local equilibrium. It means that the system may be divided into small regions characterized by mean values of extensive variables, such as mass, energy and momentum. The scalar mass and energy and the vector momentum make together five variables for a one-component system. They are changed in the course of the elementary processes of mass, heat and momentum transfer between regions neighboring in space. The phenomenological laws describing these processes form a closed set of equations. Turning the volume of a region to zero, one obtains a set of five hydrodynamic equations for the continuous fields of the densities of extensive variables. These equations may be solved provided the phenomenological transport coefficients (the thermal conductivity, the shear and volume viscosities) are known. The transport coefficients may be either extracted from experimental data or calculated theoretically from a microscopic model of the liquid. The latter way is possible if the system is close to the thermodynamic equilibrium and the fluctuations of the introduced collective variables are small as compared to their mean values. For a system of $n$ components, the set of hydrodynamic equations consists of $5n$ equations involving some transport coefficients accounting for coupling between all the components.

The hydrodynamic approach described above works fairly good for systems with a simple molecular structure. In fact, its success depends on the applicability of the phenomenological equations for the transport processes between elementary regions. It implies, first, that the size of an elementary region is large in comparison with the free path of a molecule in the system. Second, any relaxation process within an elementary volume should take much less time than the hydrodynamic relaxation of fluctuations of the collective variables. For simple liquids, this sets the limit for the hydrodynamic equations to be valid only at scales greater than ~10Å.

A polymeric liquid, even consisting of relatively short molecules obeying the Rouse dynamics [6], reveals a broad spectrum of relaxation times. Thus, the usual hydrodynamic approach would be validated only at scales exceeding the size of a polymer coil and its largest characteristic time. Does it mean that the relaxation within a vast space-time region corresponding to the internal processes of a polymer coil cannot be described in terms of the collective variables? Fortunately, it is not the case. There is a theoretical scheme known as the generalized hydrodynamics, which makes it possible to overcome the aforementioned restrictions. (Its presentation for low-molecular liquids with anomalous properties like high viscosity is given, e.g., in [34].) In this section we give a short but concise description of this scheme enabling us to describe the relaxation of coupled number density, energy density, and density flux fluctuations in multi-component polymer systems.

We consider a compressible homopolymer blend of $L$ components specified by the monomer units positions $\mathbf{r}_{il}$ and velocities $\mathbf{v}_{il}$, where $l$ numerates the units of the $i$-th kind, $1 \leq i \leq L$. We characterize the state of the blend by the distributions of the local number density $\rho_i(\mathbf{r}) = \sum_l \delta(\mathbf{r} - \mathbf{r}_{il})$, the quantity $\theta_i(\mathbf{r}) = m_i \sum_l (v_{il}^2/3) \delta(\mathbf{r} - \mathbf{r}_{il}) = \rho_i(\mathbf{r}) T_i(\mathbf{r})$, which is proportional to the kinetic energy density, and the density flux $\mathbf{j}_i(\mathbf{r}) = \sum_l \mathbf{v}_{il} \delta(\mathbf{r} - \mathbf{r}_{il})$. Here the summations are performed over all units of the $i$-th kind, $T_i(\mathbf{r})$ being the local temperature of the $i$-th component measured in the energy units.

We suppose further that the system is in a state of partial thermodynamic equilibrium characterized by the macroscopic variables $\bar{\rho}_i(\mathbf{r}) = \langle \rho_i(\mathbf{r}) \rangle$, $\bar{\theta}_i(\mathbf{r}) = \langle \theta_i(\mathbf{r}) \rangle$, $\bar{\mathbf{j}}_i(\mathbf{r}) = \langle \mathbf{j}_i(\mathbf{r}) \rangle$, where averaging is assumed to be over all microscopic states corresponding to the state of partial equilibrium. In what follows we treat all these $5 \times L$-variables as the components of one vector $\vec{R}_i(\mathbf{r}) = \{\rho_i(\mathbf{r}), \theta_i(\mathbf{r}), \mathbf{j}_i(\mathbf{r})\}$ (the 5d- and 3d-vectors are designated by the arrowed and bold letters, respectively) and use, for brevity, the matrix form of the corresponding equations.

In the state of complete thermodynamic equilibrium all the fluxes vanish ($\bar{\mathbf{j}}_i(\mathbf{r}) = 0$) and equilibrium temperatures equal the same constant value ($\overline{T}_i(\mathbf{r}) = T$). Thus, the inequilibrium states with non-zero fluxes and different temperatures of the components could exist only if some external forces drive the system away from the equilibrium state. If these forces are small, then the dynamics of the system may be described within the Onsager linear theory as a response to these forces:

$$\vec{\Delta}_i(\mathbf{r},t) = -\int_0^\infty d\tau \int d\mathbf{r}' \hat{\alpha}_{ik}(\mathbf{r}-\mathbf{r}',\tau)\vec{\varepsilon}_k(\mathbf{r}',t-\tau). \qquad (2.1a)$$

where $\vec{\Delta}_i(\mathbf{r}) = \{\rho_i(\mathbf{r}) - \overline{\rho}_i(\mathbf{r}), \theta_i(\mathbf{r}) - \overline{\theta}_i(\mathbf{r}), \mathbf{j}_i(\mathbf{r})\} = \{\delta\rho_i(\mathbf{r}), \delta\theta_i(\mathbf{r}), \mathbf{j}_i(\mathbf{r})\}$ is the deviation of the system from equilibrium.

Applying the Fourier-Laplace transform $f(\mathbf{q},p) = \int_0^\infty dt \int d\mathbf{r}\, f(\mathbf{r},t)\exp(i\mathbf{q}\mathbf{r} - pt)$ to both sides of Eq. (2.1a), we reduce it to the simple form

$$\vec{\Delta}_i(\mathbf{q},p) = -\hat{\alpha}_{ik}(\mathbf{q},p)\vec{\varepsilon}_k(\mathbf{q},p), \qquad (2.1b)$$

where $\vec{\varepsilon}_k(\mathbf{r},t) = \{\varepsilon_k^\rho(\mathbf{r},t), \varepsilon_k^\theta(\mathbf{r},t), \varepsilon_k^{\mathbf{j}}(\mathbf{r},t)\}$ is an external $5\times L$-component field applied to the system and $\hat{\alpha}_{ik}(\mathbf{q},p)$ are $5\times 5$-submatrices of the $(5\times L)\times(5\times L)$-matrix of the generalized susceptibility, $\hat{\boldsymbol{\alpha}}(\mathbf{q},p) = \|\hat{\alpha}_{ik}(\mathbf{q},p)\|$.

Hereafter, we refer to functions and their Fourier-Laplace transforms as the same functions in ($\mathbf{r}$,$t$)- and ($\mathbf{q}$,$p$)-representation, respectively, and distinguish them only by the choice of the letters used to denote their arguments; the summation convention will also be in force for repeated indices, so that $a_i b_i$ denotes a sum over $i$ from 1 to $L$; the subscripts of the components of the vector $\vec{\varepsilon}_k$ indicate the sort of the monomer units and superscripts do the type of the local variable characterizing this sort; the $(5\times L)\times(5\times L)$-matrices and their $5\times 5$-submatrices are denoted by the bold capped and capped letters, respectively.

Thus, Eq. (2.1) is a natural extension of the generalized susceptibility definition used in the conventional dynamic RPA [19,20] to describe the pure diffusion only. According to the fluctuation-dissipation theorem [16,17], the generalized susceptibility $\hat{\boldsymbol{\alpha}}(\mathbf{q},t)$ and structure factor of a system

$$\hat{\mathbf{S}}(\mathbf{q},t) = \|\hat{S}_{ik}(\mathbf{q},t)\| = \|\langle\vec{\Delta}_i(\mathbf{q},t)\vec{\Delta}_k(-\mathbf{q},0)\rangle\|/V, \qquad (2.2)$$

where V is the volume of the system, satisfy the relation

$$\hat{\boldsymbol{\alpha}}(\mathbf{q},t) = -\frac{\partial}{\partial t}\hat{\mathbf{S}}(\mathbf{q},t), \qquad (2.3)$$

which in the Fourier-Laplace representation reads

$$T\hat{\mathbf{\alpha}}(\mathbf{q}, p) = \hat{\mathbf{G}}(\mathbf{q}) - p\hat{\mathbf{S}}(\mathbf{q}, p), \tag{2.4}$$

where $\hat{\mathbf{S}}(\mathbf{q}, p) = \int_0^\infty dt \hat{\mathbf{S}}(\mathbf{q},t) \exp(-pt)$ and $\hat{\mathbf{S}}(\mathbf{q},t)\big|_{t=0} = \hat{\mathbf{G}}(\mathbf{q})$.

On the other hand, relaxation of the components of the vector $\vec{\Delta}_i(\mathbf{q},t)$ is described by the set of linearized hydrodynamic equations

$$\frac{\partial \vec{\Delta}_i(\mathbf{q},t)}{\partial t} = -q^2 \int_0^\infty d\tau \hat{\Lambda}_{ik}(\mathbf{q},\tau)[\delta\vec{\mu}_k(\mathbf{q},t-\tau) + \vec{\varepsilon}_k(\mathbf{q},t-\tau)]/T, \tag{2.5a}$$

or

$$p\vec{\Delta}_i(\mathbf{q},p) + q^2 \hat{\Lambda}_{ik}(\mathbf{q},p)[\delta\vec{\mu}_k(\mathbf{q},p) + \vec{\varepsilon}_k(\mathbf{q},p)]/T = \vec{\Delta}_i(\mathbf{q},t)\big|_{t=0} \tag{2.5b}$$

Here $\hat{\mathbf{\Lambda}}(\mathbf{q},p) = \|\hat{\Lambda}_{ik}(\mathbf{q},p)\|$ is the generalized kinetic coefficient matrix; the generalized thermodynamic force $\delta\vec{\mu}_k(\mathbf{r},t) = \delta F(\{\vec{\Delta}_i(\mathbf{r},t)\})/\delta\vec{\Delta}_k(\mathbf{r},t)$ is the variational derivative with respect to the local deviation from the equilibrium $\vec{\Delta}_i(\mathbf{r},t)$ of the free energy of a weakly non-equilibrium system $\Delta F(\{\vec{\Delta}_i(\mathbf{r},t)\})$. In the linear theory

$$\delta\vec{\mu}_k(\mathbf{q},t) = T(\hat{\mathbf{G}}^{-1}(\mathbf{q}))_{kl} \vec{\Delta}_l(\mathbf{q},t) \tag{2.6}$$

where

$$\hat{\mathbf{G}}(\mathbf{q}) = \|\hat{G}_{ik}(\mathbf{q})\| = \|\langle\vec{\Delta}_i(\mathbf{q},t)\vec{\Delta}_k(-\mathbf{q},t)\rangle_{t=0}\|/V \tag{2.7}$$

is the $(5 \times L) \times (5 \times L)$ matrix of the static correlation functions.

Eqs. (2.5) are referred to as the generalized hydrodynamics equations. They preserve the formal structure of the conventional (linearized) hydrodynamic equations except that now the transport coefficients reveal a considerable dependence on the wave number $q$ and the Laplace transform variable $p$ (or frequency $w = ip$). This dependence is referred to as the wave number and frequency dispersion and it keeps the whole information about the non-local relaxation and dissipation behavior of the system under consideration.

Indeed, let a finite deviation from equilibrium $\vec{\Delta}_i(\mathbf{q})$ be stabilized during a long (in fact, infinite) time by the corresponding external field and the field switched out at the moment $t = 0$. It follows from Eq. (2.5) that in this case the subsequent relaxation of this deviation in the $(\mathbf{q},p)$-representation obeys the relationship

$$\left(p\hat{E}_{ik} + q^2 \hat{D}_{ik}(\mathbf{q},p)\right)\vec{\Delta}_k(\mathbf{q},p) = \vec{\Delta}_i(\mathbf{q}), \tag{2.8}$$

where $\hat{\mathbf{E}}$ is the identity matrix and

$$\hat{\mathbf{D}}(\mathbf{q}, p) = \hat{\mathbf{\Lambda}}(\mathbf{q}, p)\hat{\mathbf{G}}^{-1}(\mathbf{q}) \qquad (2.9)$$

is the matrix of the generalized diffusion coefficients.

Comparing Eqs. (2.1) and (2.5), it easy to see that the matrix $\hat{\boldsymbol{\alpha}}$ is related to the matrices $\hat{\mathbf{G}}$ and $\hat{\mathbf{\Lambda}}$ by

$$\hat{\boldsymbol{\alpha}}^{-1}(\mathbf{q}, p) = T\left(\hat{\mathbf{G}}^{-1}(\mathbf{q}) + \left(p/q^2\right)\hat{\mathbf{\Lambda}}^{-1}(\mathbf{q}, p)\right) \qquad (2.10)$$

Thus, the key problem within the approach we outlined here is to evaluate the transport coefficients (or susceptibilities, or time-correlation functions) matrices with due regard for their temporal and spatial dispersion.

The presented combination of all the transfer processes into one transfer process of a $5 \times L$-vector quantity could appear to be rather formal. In particular, to define the fields driving the system under consideration away from its equilibrium state is easy only for the components $\{\varepsilon_k^\rho(\mathbf{r},t)\}$. To do it for the components $\{\varepsilon_k^\theta(\mathbf{r},t), \varepsilon_k^j(\mathbf{r},t)\}$ is an old and hard problem (see, e.g., [35]). But what we will really need is the susceptibility matrix $\hat{\boldsymbol{\alpha}}$ rather than the fields $\{\varepsilon_k^\theta(\mathbf{r},t), \varepsilon_k^j(\mathbf{r},t)\}$.

It could seem hopeless to calculate the generalized susceptibility directly for a system with a specified polymer structure and volume interactions (by volume interaction, we mean the interaction that remains after all the chemical bonds forming the polymer system have been broken). Nevertheless, considerable progress can be made along the line, which was first proposed by I. Lifshitz [36] and is as follows. We assume that all the relevant properties are already known for a system of small molecules we refer to as the broken links system (BLS), which models the volume interaction for our polymer system. The problem is then to express the polymer dynamic properties in terms of the BLS properties and those of some model polymer systems. In the next section we demonstrate this approach in action.

# 3 The generalized dynamic RPA and the structural and direct susceptibilities

It is to find the generalized susceptibility $\hat{\alpha}$, which is the primary goal of the DRPA. In this section we extend the derivation of this approximation given in [19] to the case when the fluctuations $\vec{\Delta}_i(\mathbf{r},t)$ of the number and energy densities as well as the density flux are coupled. The starting point of this derivation is the simple fact that the collective response (2.1) of the polymer system to an external field is, within the self-consistent field (SCF) approximation, just the sum of the responses of the individual macromolecules:

$$\vec{\Delta}_i(\mathbf{r},t) = \sum_{\{M\}} \vec{\Delta}_i^{(M)}(\mathbf{r},t,\{\varepsilon_M(\Gamma)\}), \tag{3.1}$$

where the summation is over all sorts of different macromolecules in the system. The form of (3.1) indicates that the partial contribution $\vec{\Delta}_i^{(M)}$ from macromolecules $M$ is a functional that depends on an effective field $\vec{\varepsilon}_M(\Gamma)$, $\Gamma$ being a point in the phase space of $M$. In general, $\vec{\varepsilon}_M(\Gamma)$ depends on the structure of $M$ and is *not* equal to the sum of the external fields acting on the structural units of the macromolecules of the sort $M$. The case is that the external fields drive the system away from its state of thermodynamic equilibrium, so that the average interaction energy among the linked monomers of $M$ and between them and the rest of the system changes accordingly.

Now, it seems reasonable to suppose that in the dense systems the effective field $\vec{\varepsilon}_M(\Gamma)$ is basically determined by the inter-chain interactions, whereas the contribution of the intra-chain interaction to $\vec{\varepsilon}_M(\Gamma)$ can be neglected to a first approximation. Then we have

$$\vec{\varepsilon}_M(\Gamma) = \sum \vec{\varepsilon}_i^{eff}(\mathbf{r}_{l_i},t) = \sum \{\vec{\varepsilon}_i^e(\mathbf{r}_{l_i},t) + \vec{\varepsilon}_i^*(\mathbf{r}_{l_i},t)\}. \tag{3.2}$$

Here the sum includes all the monomer units belonging to all macromolecules of the sort $M$, and the effective fields $\vec{\varepsilon}_i^{eff}$ differ from the "bare" external fields $\vec{\varepsilon}_i^e$ by the "molecular" fields $\vec{\varepsilon}_i^*$ to be obtained by a proper averaging the interaction between the $l$-th monomer unit of type $i$ and all the monomer units that surround it *and* belong to other macromolecules.

Since such an averaging is carried out over the scales comparable to "low-molecular" characteristic scales, the result should be insensitive to the polymer structure of the system and, therefore, it should be the same as for the broken links system. Within the linear theory we restrict ourselves here, this molecular fields read

$$\vec{\varepsilon}_i^*(\mathbf{r},t) = -T\int_0^\infty d\tau \int d\mathbf{r}' \hat{d}_{ij}(\mathbf{r}-\mathbf{r}',\tau)\vec{\Delta}_j(\mathbf{r}',t-\tau) \tag{3.3a}$$

in ($\mathbf{r}$,$t$)-representation or

$$\vec{\varepsilon}_i^*(\mathbf{q},p) = -T\,\hat{d}_{ik}(\mathbf{q},p)\vec{\Delta}_k(\mathbf{q},p) \tag{3.3b}$$

in (**q**,*p*)-representation. We refer to the matrix $\hat{\mathbf{d}}$, characterizing the non-equilibrium behavior of the BLS close to equilibrium, as the direct susceptibility.

The partial contribution from macromolecule *M* to the collective response of the system is

$$-\vec{\Delta}_i^{(M)}(\mathbf{q},p) = \hat{\gamma}_{ik}^{(M)}(q,p)\vec{\varepsilon}_k^{\text{eff}}(\mathbf{q},p) \tag{3.4}$$

where, by analogy with (2.4), the molecular susceptibility $\hat{\gamma}^{(M)}$ of *M* reads [19]

$$T\hat{\gamma}^{(M)}(\mathbf{q},p) = \hat{\Gamma}^{(M)}(\mathbf{q}) - p\int_0^\infty dt\,\hat{\sigma}^{(M)}(\mathbf{q},t)\exp(-pt). \tag{3.5}$$

Here

$$\hat{\sigma}_{ik}^{(M)}(\mathbf{q},t) = \int d\mathbf{r}\,\left\langle \vec{\Delta}_i(\mathbf{r},t)\vec{\Delta}_k(0,0)\right\rangle_M \exp(i\mathbf{q}\mathbf{r}),\quad \hat{\Gamma}^{(M)}(\mathbf{q}) = \hat{\sigma}^{(M)}(\mathbf{q},t)\big|_{t=0} \tag{3.6}$$

are, respectively, the generalized dynamic and static structure factors corresponding to the ensemble of all chains of the sort *M* we calculate in the next section.

Now, substituting into the left and right sides of Eq. (3.1) the value of the collective response (2.1) and the sum over all molecular responses as consistent with Eqs. (3.2), (3.3) and (3.4), we get finally the desired relation

$$\hat{\boldsymbol{\alpha}}^{-1}(\mathbf{q},p) = T(\hat{\gamma}^{-1}(\mathbf{q},p) - \hat{\mathbf{d}}(\mathbf{q},p)), \tag{3.7}$$

where the quantity

$$\hat{\gamma}(\mathbf{q},p) = T\sum_{\{M\}}\hat{\gamma}^{(M)}(\mathbf{q},p) \tag{3.8}$$

is referred to as the structural susceptibility matrix.

The relation (3.7) is the central result of the generalized dynamic RPA we are developing and it deserves some discussion. First, it means that calculation of the generalized susceptibility $\hat{\gamma}$ characterizing the collective linear relaxation via coupled mass, energy and flux transfer processes of a dense polymer system (generally, polydisperse and heteropolymer) is reduced, within the DRPA, to that of two simpler quantities: the structural and direct susceptibilities. Indeed, just by construction, the structural susceptibility $\hat{\gamma}^{(M)}$ characterizes the averaged off-equilibrium (but close to equilibrium) dynamics of chain *M* subjected to an external (even though, in fact, self-consistent) field. That is why it could be calculated using a suitable dynamic model of a single chain. In particular, for the purposes of the present paper we need the structural susceptibility $\hat{\gamma}^{(M)}$ for the Rouse model, which we calculate in section 4 and in the Appendix 1. On the other hand, we circumvent the problem of taking into account the presence of strong interaction between the monomer units (just because it is strong!) by introducing the direct susceptibility $\hat{\mathbf{d}}$ we calculate in section 5 comparing the conventional hydrodynamic and DRPA descriptions of simple liquids.

Second, the fact that, according to (3.7), the matrix $\hat{\boldsymbol{\alpha}}^{-1}$ is a sum of a structure-entropy (the inverse structural susceptibility) and energy (the direct susceptibility taken with minus) contributions has a very clear physical meaning. Indeed, according to the fluctuation-dissipation theorem [16,17], the inverse susceptibility is just the correlator of the random forces (heat noise) causing the dynamic fluctuations of the system. Thus, the aforementioned additivity implies that the random forces causing *i*) the elementary "jumping" processes of monomers belonging to the same macromolecule and resulting in its conformational fluctuations and displacements as the whole and *ii*) the collective "elementary jumps" of those belonging to different macromolecules are statistically independent. This additivity is, of course, only an approximation but rather natural one since the random forces assumed to be statistically independent are related to rather different space-time correlation scales.

It is worth also to mention that the idea to use the Lifshitz approach to separate the long-range (polymer structure induced) and short-range (interaction induced) correlations in polymer systems was first elaborated by one of us [37] to obtain the most general RPA expression for the $M \times M$-matrix of the static density-density correlation functions in polydisperse heteropolymer systems:

$$\mathbf{G}^{-1}(\mathbf{q}) = \mathbf{g}^{-1}(\mathbf{q}) - \mathbf{c}(\mathbf{q}), \tag{3.9}$$

In expression (3.9), which was found also in refs. [38-41], $\mathbf{c}(\mathbf{q}) = \mathbf{d}(\mathbf{q}, p = 0)$ is the matrix of the direct correlation functions well-known in the theory of simple liquids [42] and the matrix $\mathbf{g}(\mathbf{q}) = \sum_{\{M\}} \boldsymbol{\Gamma}^{(M)}(\mathbf{q})$ is referred to as the structural one, which explains the choice of the names for the corresponding susceptibilities.

The expression (3.9) was extended to $L \times L$-matrices of the density-density time-correlation functions and the corresponding susceptibilities in refs [18-22]. The expression (3.7) for the $(5 \times L) \times (5 \times L)$-matrix of the generalized susceptibility for the coupled hydrodynamic relaxation of the masses, energies and fluxes fluctuations we first presented in this paper seems to be the utmost extension along this vein.

# 4 Calculation of the structural susceptibility $\hat{\gamma}$

The basic tool we use to calculate the structural susceptibility is the formula (3.5) relating the molecular susceptibility and structure factors. Using this formula we can find the desired structural susceptibility if we calculate the single-chain structure factors (3.6) for a chosen model of polymer dynamics. Within the original DRPA such a procedure was demonstrated in refs [18-27, 31-33] when working with the dynamic structure factor $\sigma_{ik}^{\rho\rho}(\mathbf{r},t) = \langle \delta\rho_i(\mathbf{r},t)\delta\rho_k(0,0)\rangle$. But now our description includes the simultaneous relaxation of the number and energy densities as well as density fluxes, which implies using the two-dimensional $(5\times L)\times(5\times L)$-matrices $\hat{\sigma}^{(M)}$ describing the time-correlations of all these quantities for macromolecules $M$.

In the Fourier representation, the elements of the dynamic correlation functions $\hat{\sigma}^{(M)}$ have the form

$$\left(\sigma^{(M)}\right)_{ik}^{ab}(\mathbf{q},t) = \nu_M \sum_{\{n_i,l_k\}} \left\langle \begin{pmatrix} 1 & m_k v_{l_k}^2(0)/3 \\ m_i v_{n_i}^2(t)/3 & \left(m_i v_{n_i}^2(t)/3\right)\left(m_k v_{l_k}^2(0)/3\right) \end{pmatrix} \exp\left(i\mathbf{q}\left(\mathbf{r}_{n_i}(t) - \mathbf{r}_{l_k}(0)\right)\right) \right\rangle_M, \qquad (4.1a)$$

$$\left(\sigma^{(M)}\right)_{ik}^{a\alpha}(\mathbf{q},t) = \nu_M \sum_{\{n_i,l_k\}} \left\langle \begin{pmatrix} 1 \\ m_i v_{n_i}^2(t)/3 \end{pmatrix} v_{n_i}^\alpha(t) \exp\left(i\mathbf{q}\left(\mathbf{r}_{n_i}(t) - \mathbf{r}_{l_k}(0)\right)\right) \right\rangle_M, \qquad (4.1b)$$

$$\left(\sigma^{(M)}\right)_{ik}^{\alpha\beta}(\mathbf{q},t) = \nu_M \sum_{\{n_i,l_k\}} \left\langle v_{n_i}^\alpha(t) v_{l_k}^\beta(0) \exp\left(i\mathbf{q}\left(\mathbf{r}_{n_i}(t) - \mathbf{r}_{l_k}(0)\right)\right) \right\rangle_M, \qquad (4.1c)$$

where $\nu_M$ is the number of the macromolecules $M$ per unit volume, the indices $i$ and $k$ label the sorts of different segments (of mass $m_i$ and $m_k$, respectively) that belong to the chosen macromolecule and summation is performed over all numbers $n_i$ and $l_k$ of the segments of the two chosen sorts. The Roman indices $a,b = \{\rho,\theta\}$ are related to the first two components of the vector $\vec{\Delta}_i$ (fluctuations of the number density $\rho_i$ and energy density $\theta_i$), while the Greek indices $\alpha,\beta = \{x,y,z\}$ denote the coordinate axes onto which the segments velocities are projected. The designation $\langle\ \rangle_M$ implies statistical averaging over all dynamical conformations of a macromolecule $M$.

Due to isotropy of the systems we are considering in this paper the components of the tensor (4.1) are closely interrelated. To begin with, the components $\left(\sigma^{(M)}\right)_{ik}^{a\alpha}(\mathbf{q},t)$ of this tensor can be considered as some 3d vectors, which should be proportional to the only vector $\mathbf{q}$ breaking isotropy:

$$\left(\sigma^{(M)}\right)_{ik}^{a\alpha} = -\left(iq_\alpha/q^2\right)A_{ik}^a. \tag{4.2a}$$

On the other hand, it follows from the definitions (4.1) that

$$iq_\alpha\left(\sigma^{(M)}\right)_{ik}^{a\alpha} = A_{ik}^a = \partial\left(\sigma^{(M)}\right)_{ik}^{a\rho}/\partial t. \tag{4.3a}$$

Hence,

$$\left(\sigma^{(M)}\right)_{ik}^{a\alpha} = \left(-iq_\alpha/q^2\right)\partial\left(\sigma^{(M)}\right)_{ik}^{a\rho}/\partial t. \tag{4.4a}$$

Similarly, the components $\left(\sigma^{(M)}\right)_{ik}^{\alpha\beta}(\mathbf{q},t)$ of the tensor (4.1) are the tensors of the rank 2 that are to be of the form

$$\left(\sigma^{(M)}\right)_{ik}^{\alpha\beta} = -\left(q_\alpha q_\beta/q^4\right)B_{ik} + \left(\delta_{\alpha\beta} - \left(q_\alpha q_\beta/q^2\right)\right)\left(H_{ik}/q^2\right) \tag{4.2b}$$

and obey the equalities

$$\begin{aligned}iq_\alpha iq_\beta\left(\sigma^{(M)}\right)_{ik}^{\alpha\beta} &= B_{ik} = \partial^2\left(\sigma^{(M)}\right)_{ik}^{\rho\rho}/\partial t^2,\\ Sp\left(\sigma^{(M)}\right)_{ik}^{\alpha\beta} &= \sum_{\alpha=1}^{3}\left(\sigma^{(M)}\right)_{ik}^{\alpha\alpha} = (2H_{ik} - B_{ik})/q^2.\end{aligned} \tag{4.3b}$$

Thus, we get

$$\left(\sigma^{(M)}\right)_{ik}^{\alpha\beta} = -\left(q_\alpha q_\beta/q^4\right)\left(\partial^2\left(\sigma^{(M)}\right)_{ik}^{\rho\rho}/dt^2\right) + \left(\delta_{\alpha\beta} - \left(q_\alpha q_\beta/q^2\right)\right)\left(H_{ik}/q^2\right) \tag{4.4b}$$

where

$$H_{ik}(\mathbf{q},t) = \left(\left(\partial^2\left(\sigma^{(M)}\right)_{ik}^{\rho\rho}/dt^2\right) + q^2\, Sp\left(\sigma^{(M)}\right)_{ik}^{\alpha\beta}\right)/2. \tag{4.4c}$$

Taking into account Eq. (3.5) for the structural susceptibility $\hat{\gamma}$ and the fact that the one-time fluctuations $\delta\rho(0,t)$ and $\delta\mathbf{v}(\mathbf{r},t)$ are not correlated, and, hence, for any $\alpha$

$$\left(\sigma^{(M)}\right)_{ik}^{\rho\alpha}(\mathbf{q},t)\Big|_{t=+0} = \left(\partial\left(\sigma^{(M)}\right)_{ik}^{\rho\rho}(\mathbf{q},t)/\partial t\right)\Big|_{t=+0} = 0, \tag{4.5}$$

it is easy to see that in the Fourier-Laplace representation Eqs. (4.4) read

$$\left(\sigma^{(M)}\right)_{ik}^{a\alpha}(\mathbf{q},p) = \left(-iq_\alpha/q^2\right)\left(\left(\sigma^{(M)}\right)_{ik}^{a\rho}(\mathbf{q},t)\Big|_{t=+0} - p\left(\sigma^{(M)}\right)_{ik}^{a\rho}(\mathbf{q},p)\right) = \left(-iq_\alpha/q^2\right)T\left(\gamma^{(M)}\right)_{ik}^{a\rho}, \tag{4.6a}$$

$$\begin{aligned}\left(\sigma^{(M)}\right)_{ik}^{\alpha\beta}(\mathbf{q},p) &= -\frac{q_\alpha q_\beta}{q^4}\left(p^2\left(\sigma^{(M)}\right)_{ik}^{\rho\rho}(\mathbf{q},p) - p\left(\sigma^{(M)}\right)_{ik}^{\rho\rho}(\mathbf{q},t)\Big|_{t=+0} - \partial\left(\sigma^{(M)}\right)_{ik}^{\rho\rho}(\mathbf{q},t)/\partial t\Big|_{t=+0}\right)\\ &+ \left(\delta_{\alpha\beta} - \frac{q_\alpha q_\beta}{q^2}\right)\frac{H_{ik}(q,p)}{q^2} = \frac{q_\alpha q_\beta}{q^4}p\left(\gamma^{(M)}\right)_{ik}^{\rho\rho}(\mathbf{q},p) + \left(\delta_{\alpha\beta} - \frac{q_\alpha q_\beta}{q^2}\right)\frac{H_{ik}(q,p)}{q^2}\end{aligned} \tag{4.6b}$$

It follows from Eqs. (3.5), (4.5) and (4.6) that $\hat{\gamma}^{(M)}$ takes the following comparatively simple form:

$$\hat{\gamma}^{(M)}(\mathbf{q},p) = \begin{pmatrix} \left(\gamma^{(M)}\right)_{ik}^{ab} & -\dfrac{iq_\beta p}{q^2}\left(\gamma^{(M)}\right)_{ik}^{a\rho} \\ -\dfrac{iq_\alpha p}{q^2}\left(\gamma^{(M)}\right)_{ik}^{\rho b} & \Gamma_{ik}^{\alpha\beta} - \dfrac{q_\alpha q_\beta p^2}{q^4}\left(\gamma^{(M)}\right)_{ik}^{\rho\rho} \end{pmatrix}, \qquad (4.7)$$

where

$$T\Gamma_{ik}^{\alpha\beta} = \left(\sigma^{(M)}\right)_{ik}^{\alpha\beta}(\mathbf{q},t)\Big|_{t=+0} - \left(p/q^2\right)H_{ik}(\mathbf{q},p)\left(\delta_{\alpha\beta} - q_\alpha q_\beta/q^2\right). \qquad (4.8)$$

We remember that the correlations between different chains are not considered at this stage so that the structural susceptibility is calculated for each macromolecule separately. To obtain the matrix $\hat{\gamma}$ for the whole system, we should sum the contributions from all chains as consistent with the definition (3.8).

Now we proceed to explicit calculation of the structural susceptibility matrix for a disentangled (Rouse) monodisperse homopolymer melt. Let $\rho$, $\rho/N$ and $N$ be the total numbers of segments and chains per unit volume and the number of segments per chain, respectively.

First of all, we calculate the dynamical structural function

$$\sigma_{\rho\rho}(\mathbf{q},t) = (\rho/N)\sum_{k,n}\langle\exp(i\mathbf{q}(\mathbf{r}_k(t)-\mathbf{r}_n(0)))\rangle, \qquad (4.9)$$

with $k,n$ enumerating segments of a chosen chain. The well-known representation of $S_{\rho\rho}$ in the form of asymptotic series in Rouse modes [6] is inconvenient to us since it cannot be directly subjected to the Laplace transformation. To avoid this difficulty, instead of solving the Langevin equations for the radius-vectors of chain segments and substituting the obtained expression into (4.9), we derived an equation directly for the correlator $\langle\exp(i\mathbf{q}(\mathbf{r}_k(t)-\mathbf{r}_n(0)))\rangle$ in the Rouse model (see Appendix 1).

This equation turns out to be the diffusion-like one and may be easily solved by the Laplace transformation as it was done for the reptation model [19]. Then, calculating the structure factor and using Eq. (3.5), we get the desired element of the structural susceptibility matrix $\gamma_{\rho\rho}$ in the form

$$\gamma_{\rho\rho}(\mathbf{q},p) = \frac{\rho N Q^2}{Q^2 - K}\left(\frac{1}{K + Q\sqrt{K}\coth\sqrt{K}} - \frac{1}{Q^2(1+\coth Q)}\right), \qquad (4.10)$$

Here we introduced the "polymeric" dimensionless variables $Q = q^2R^2/2$ and $K = p\tau_R/2$, where $R^2 = Nb^2/6$ is the chain mean-square radius, $\tau_R = N^2b^2/(6D_0)$ is the time needed for the chain center of mass to diffuse over distance $R$, $b$ is the length of the statistical segment and $D_0$ is the segment diffusivity for the Rouse model. Thus, the self-diffusion coefficient of the chain is $D = R^2/\tau_R = D_0/N$. Note that $\tau_R$ and the maximum relaxation time of the Rouse spectrum (Rouse time) $\tau_1 = N^2 b^2/(3\pi^2 D_0)$ [6] differ in a numerical factor only.

It is convenient also to give here the expression for the inverse quantity we use in Section 6:

$$\frac{\rho N}{\gamma_{\rho\rho}(Q,K)} = \frac{1}{f_D(2Q)} + \frac{K}{Q}\Lambda^{-1}(Q,K), \qquad (4.11a)$$

where $f_D(x) = 2(\exp(-x) - 1 + x)/x^2$ is the Debye function and the inverse kinetic coefficient reads [18]:

$$\Lambda^{-1}(Q,K) = \lambda_\infty^{-1}(Q) + F(Q,K), \qquad \lambda_\infty^{-1}(Q) = Q(1 + \coth Q),$$
$$F(Q,K) = \frac{\lambda_\infty^{-1}(Q) f_D^{-1}(Q)}{K} \frac{Q^2(\sqrt{K}\coth\sqrt{K} - 1) - K(Q\coth Q - 1)}{Q^2(1 + \coth Q) - (K + \sqrt{K}\coth\sqrt{K})} \qquad (4.11b)$$

The high-frequency limit $\lambda_\infty(Q)$ of the kinetic coefficient and the correction $F(Q,P)$ were first calculated for the reptation model in Refs [43] and [19], respectively.

To obtain other elements of the matrix $\hat{\gamma}$, we are to calculate the correlators (4.1) that contain both the density of chain segments and their velocity. Within the framework of the Rouse model, the components of the segment displacements $(r_k^\alpha(t) - r_n^\alpha(0))$ and velocities $v_k^\alpha$ are random Gaussian variables. This reduces the problem to calculation of their average values and dispersions [6]. But finding of the asymptotic large-scale (as compared to the microscopic size $b$ and time $\tau_0 = b^2/D_0$) behavior is rather simple. Since the heat segment velocity equilibrates much faster than the chain conformations, we just have

$$\sigma_{\rho\theta}(\mathbf{q},t) = m(\rho/N)\sum_{k,l}\langle(v_k^2(t)/3)\exp(i\mathbf{q}(\mathbf{r}_k(t) - \mathbf{r}_l(0)))\rangle = T\sigma_{\rho\rho}(\mathbf{q},t), \qquad (4.12a)$$

$$\sigma_{\rho\theta}(\mathbf{q},0) = m(\rho/N)\sum_{k,l}\langle(v_k^2(0)/3)\exp(i\mathbf{q}(\mathbf{r}_k(0) - \mathbf{r}_l(0)))\rangle = T\sigma_{\rho\rho}(\mathbf{q},0). \qquad (4.12b)$$

where $m$ is the segment mass. Similarly,

$$\sigma_{\theta\theta}(\mathbf{q},t) = (\rho/N)\sum_{k,l}\langle(mv_k^2(t)/3)(mv_l^2(0)/3)\exp(i\mathbf{q}(\mathbf{r}_k(t) - \mathbf{r}_l(0)))\rangle = T^2\sigma_{\rho\rho}(\mathbf{q},t), \qquad (4.13a)$$

$$\sigma_{\theta\theta}(\mathbf{q},0) = (\rho/N)\sum_{k,l}\langle(mv_k^2(0)/3)(mv_l^2(0)/3)\exp(i\mathbf{q}(\mathbf{r}_k(t) - \mathbf{r}_l(0)))\rangle$$
$$= T^2\sigma_{\rho\rho}(\mathbf{q},0) + (2/3)T^2\rho, \qquad (4.13b)$$

where the last term is the correction due to the fact that for $k=l$ the density-density correlation function should be multiplied by $\langle v^4/9 \rangle$ rather than .

At last, the one-time correlation function appearing in Eq. (4.8) for homopolymer melt is

simply

$$\sigma^{\alpha\beta}(\mathbf{q},t)\big|_{t=0} = (\rho/N)\sum_{\{n,l\}}\langle v_n^\alpha(0) v_l^\beta(0) \exp(i\mathbf{q}(\mathbf{r}_n(0) - \mathbf{r}_l(0)))\rangle = \delta_{\alpha\beta}\Gamma(q).$$

The explicit form of the function $\Gamma(q)$ does not influence the sound propagation as shown below. Collecting all the calculated correlators we finally obtain the generalized structural susceptibility of a Rouse monodisperse homopolymer melt:

$$\hat{\gamma}(\mathbf{q},p) = \begin{pmatrix} \gamma_{\rho\rho}(q,p) & T\gamma_{\rho\rho}(q,p) & -\dfrac{iq_\beta p}{q^2}\gamma_{\rho\rho}(q,p) \\ T\gamma_{\rho\rho}(q,p) & T^2\gamma_{\rho\rho}(q,p) + \dfrac{2}{3}T^2\rho & -\dfrac{iq_\beta pT}{q^2}\gamma_{\rho\rho}(q,p) \\ -\dfrac{iq_\alpha p}{q^2}\gamma_{\rho\rho}(q,p) & -\dfrac{iq_\alpha pT}{q^2}\gamma_{\rho\rho}(q,p) & \dfrac{q_\alpha q_\beta}{q^2}\left(\Gamma(q) - \dfrac{p^2}{q^2}\gamma_{\rho\rho}(q,p)\right) + \left(\delta_{\alpha\beta} - \dfrac{q_\alpha q_\beta}{q^2}\right)W \end{pmatrix}, \qquad (4.14)$$

where $W(\mathbf{q},p) = \Gamma(q) - (p/q^2)H(q,p)$ and $H(q,p)$ is the Laplace transform of the function (4.4c) for homopolymer melt.

Inverting the structural susceptibility matrix (see Appendix 2), we get

$$\hat{\gamma}^{-1}(\mathbf{q},p) = \begin{pmatrix} \dfrac{1}{\gamma_{\rho\rho}} + \dfrac{3}{2\rho} - \dfrac{p^2}{q^2\Gamma} & -\dfrac{3}{2T\rho} & \dfrac{iq_\beta p}{q^2\Gamma} \\ -\dfrac{3}{2T\rho} & \dfrac{3}{2T^2\rho} & 0 \\ \dfrac{iq_\alpha p}{q^2\Gamma} & 0 & \dfrac{q_\alpha q_\beta}{q^2}\dfrac{1}{\Gamma} + \left(\delta_{\alpha\beta} - \dfrac{q_\alpha q_\beta}{q^2}\right)\dfrac{1}{W} \end{pmatrix}. \qquad (4.15)$$

It is worth to notice that in the expressions for both $\hat{\gamma}$ and $\hat{\gamma}^{-1}$ the function $W$ appears only in the block corresponding to the flux-flux correlations.

**5 The direct susceptibility of a one-component system**

In this section we calculate the direct susceptibility $\hat{d}$ for compressible homopolymer melt, which is the simplest polymer system (with the number of components $M = 1$) suitable to demonstrate how does the generalized DRPA work. For this purpose we notice that, by definition, the general DRPA equation (3.7) is applicable also to the broken links system, where it reads

$$\hat{\alpha}_{BLS}^{-1}(\mathbf{q},p) = T(\hat{\gamma}_{BLS}^{-1}(\mathbf{q},p) - \hat{d}(\mathbf{q},p)) \qquad (5.1)$$

Here the structural susceptibility $\hat{\gamma}_{BLS}(\mathbf{q},p)$ of one-component fluid could be easily found if we set $N = 1$ in (4.1), which results in the following expressions for the corresponding self-

correlators for the BLS:

$$\sigma_{BLS}^{\rho\rho}(\mathbf{q},p) = \rho/(p+q^2 D_0), \quad \sigma_{BLS}^{\rho\theta}(\mathbf{q},p) = T\sigma_{BLS}^{\rho\rho}(\mathbf{q},p), \quad \sigma_{BLS}^{j_\alpha j_\beta}(\mathbf{q},p) = \delta_{\alpha\beta}\Gamma_0,$$
$$\sigma_{BLS}^{\rho j_\alpha}(\mathbf{q},p) = \sigma_{BLS}^{\theta j_\alpha}(\mathbf{q},p) = 0, \quad \sigma_{BLS}^{\theta\theta}(\mathbf{q},p) = T^2 \sigma_{BLS}^{\rho\rho}(\mathbf{q},p) + (2/3)\rho T^2,$$
(5.2)

where $\Gamma_0 = \rho T/m$ and $D_0$ is the self-diffusion coefficient of particles forming the BLS. The expression for the BLS structural susceptibility could be then found similarly to that for the homopolymer melt given by (4.14):

$$\hat{\gamma}_{BLS} = \begin{pmatrix} \gamma_{BLS}^{\rho\rho} & T\gamma_{BLS}^{\rho\rho} & (-ipq_\beta/q^2)\gamma_{BLS}^{\rho\rho} \\ T\gamma_{BLS}^{\rho\rho} & T^2\gamma_{BLS}^{\rho\rho} + 2\rho T^2/3 & (-ipq_\beta/q^2)T\gamma_{BLS}^{\rho\rho} \\ (-ipq_\alpha/q^2)\gamma_{BLS}^{\rho\rho} & (-ipq_\alpha/q^2)T\gamma_{BLS}^{\rho\rho} & \Gamma_0\delta_{\alpha\beta} - (p^2 q_\alpha q_\beta/q^4)\gamma_{BLS}^{\rho\rho} \end{pmatrix},$$
(5.3)

where

$$\gamma_{BLS}^{\rho\rho} = \rho q^2 D_0/(p + q^2 D_0).$$
(5.3a)

The inverse structure susceptibility for the BLS is

$$\hat{\gamma}_{BLS}^{-1}(\mathbf{q},p) = \begin{pmatrix} \dfrac{p}{q^2 D_0 \rho} + \dfrac{5}{2\rho} - \dfrac{p^2}{q^2 \Gamma_0} & -\dfrac{3}{2T\rho} & \dfrac{iq_\beta p}{q^2 \Gamma_0} \\ -\dfrac{3}{2T\rho} & \dfrac{3}{2T^2\rho} & 0 \\ \dfrac{iq_\alpha p}{q^2 \Gamma_0} & 0 & \dfrac{\delta_{\alpha\beta}}{\Gamma_0} \end{pmatrix}.$$
(5.3b)

On the other hand, the $5\times 5$-matrix $\hat{\alpha}_{BLS}$ is determined by the standard hydrodynamic description of compressible fluid and, therefore, it could be calculated independently. Then we get the direct susceptibility straightforwardly from (5.1).

**The generalized BLS susceptibility via hydrodynamics.** Let us consider the system of broken links as a simple fluid whose dynamics is described, in the linear approximation, by the conventional set of the linearized hydrodynamic equations [8]:

$$\frac{\partial \rho}{\partial t} + \text{div }\mathbf{j} = 0$$
$$\rho T \frac{\partial S}{\partial t} = \kappa \Delta T$$
$$\frac{\partial \mathbf{j}}{\partial t} = -\frac{\nabla P}{m} + \nu\Delta\mathbf{j} + \mu\nabla\text{div }\mathbf{j},$$
(5.4)

Here $\nu = \eta/(\rho m)$, $\mu = (\zeta + \eta/3)/(\rho m)$, where $\zeta$ and $\eta$ are the first and second viscosities of the BLS, $\kappa$ is the thermal conductivity, $P$ and $S$ are the local pressure and entropy per chain segment, respectively. In this paper we neglect any dispersion of the transport coefficients μ, υ, and κ and assume that there is no macroscopic flux in the system.

First, using the thermodynamic relations [16]

$$\delta S = \left(\frac{\partial S}{\partial \rho}\right)_\theta \delta\rho + \left(\frac{\partial S}{\partial \theta}\right)_\rho \delta\theta = \frac{c_V}{\rho T}\left(\delta\theta - \left(\frac{\partial \theta}{\partial \rho}\right)_S \delta\rho\right), \quad \delta P = \left(\frac{\partial P}{\partial \rho}\right)_\theta \delta\rho + \left(\frac{\partial P}{\partial \theta}\right)_\rho \delta\theta, \qquad (5.5)$$

we reduce Eqs. (5.4) to the form where the components of the fluctuation vector $\vec{\Delta}(\mathbf{r})$ only appear:

$$\frac{\partial \delta\rho}{\partial t} + \text{div}\,\mathbf{j} = 0$$

$$\frac{\partial \delta\theta}{\partial t} = -D_T T \Delta\delta\rho + D_T \Delta\delta\theta - \left(\frac{\partial \theta}{\partial \rho}\right)_S \text{div}\,\mathbf{j} \qquad (5.6)$$

$$\frac{\partial \mathbf{j}}{\partial t} = -\frac{1}{m}\left(\frac{\partial P}{\partial \rho}\right)_\theta \nabla\delta\rho - \frac{1}{m}\left(\frac{\partial P}{\partial \theta}\right)_\rho \nabla\delta\theta + \mu\nabla\text{div}\,\mathbf{j} + \nu\Delta\mathbf{j},$$

where $D_T = \kappa/(\rho c_V)$ is the thermal diffusivity, $c_V = T(\partial S/\partial T)_V = \theta(\partial S/\partial \theta)_\rho$ is the heat capacity per chain segment under constant volume and we used the fact than in the linear approximation $\Delta T = (\Delta\theta - T\Delta\rho)/\rho$.

In the Fourier representation the simultaneous linearized hydrodynamic equations (5.4) may be written in the matrix form

$$\frac{\partial \Delta_l(\mathbf{q},t)}{\partial t} = -q^2 (D_{\text{BLS}})_{lm}(\mathbf{q})\Delta_m(\mathbf{q},t), \qquad (5.7)$$

with the matrix of the generalized diffusion coefficients

$$\hat{D}_{\text{BLS}}(\mathbf{q}) = \begin{pmatrix} 0 & 0 & -iq_\beta/q^2 \\ -D_T T & D_T & -\dfrac{iq_\beta}{q^2}\left(\dfrac{\partial \theta}{\partial \rho}\right)_S \\ -\dfrac{iq_\alpha}{mq^2}\left(\dfrac{\partial P}{\partial \rho}\right)_\theta & -\dfrac{iq_\alpha}{mq^2}\left(\dfrac{\partial P}{\partial \theta}\right)_\rho & \upsilon\delta_{\alpha\beta} + \mu q_\alpha q_\beta/q^2 \end{pmatrix}. \qquad (5.8)$$

The matrix of the Onsager kinetic coefficients is

$$\hat{\Lambda}_{\text{BLS}} = \hat{D}_{\text{BLS}}\,\hat{G}_{\text{BLS}}, \qquad (5.9)$$

where the (5×5) matrix of the static correlation functions $G_{\text{BLS}}^{lm}(\mathbf{q})$ can be calculated using the general theory of thermodynamic fluctuations [16]:

$$\hat{G}_{BLS}(\mathbf{q}) = \rho T \begin{pmatrix} (\partial \rho/\partial P)_T & T(\partial \rho/\partial P)_T & 0 \\ T(\partial \rho/\partial P)_T & T^2(\partial \rho/\partial P)_T + T/c_V & 0 \\ 0 & 0 & m^{-1}\delta_{\alpha\beta} \end{pmatrix}. \qquad (5.10)$$

Using Eqs (5.6), (5.8)-(5.10) and the thermodynamic identity $(\partial S/\partial \rho)_T = -\rho^{-2}(\partial P/\partial T)_\rho$ we get

$$\hat{\Lambda}_{BLS}(\mathbf{q}) = \begin{pmatrix} 0 & 0 & -in_\beta b \\ 0 & a & -in_\beta bc \\ -in_\alpha b & -in_\alpha bc & \Gamma_0(\nu\delta_{\alpha\beta} + \mu n_\alpha n_\beta) \end{pmatrix}, \qquad (5.11)$$

where $a = \rho D_T T^2/c_V$, $b = \Gamma_0/q$, $c = (\partial \theta/\partial \rho)_S$ and $n_\alpha = q_\alpha/q$. Vanishing of the kinetic coefficients corresponding to the mass transfer is natural because there is no mass dissipation for one-component fluid.

Now, the generalized susceptibility $\hat{\alpha}_{BLS}$ is determined by the matrices $\hat{\Lambda}_{BLS}$ and $\hat{G}_{BLS}$ as consistent with the fluctuation-dissipation relation (2.10):

$$\frac{\hat{\alpha}_{BLS}^{-1}(\mathbf{q},p)}{T} = \frac{p}{q^2}\hat{\Lambda}_{BLS}^{-1}(\mathbf{q},p) + \hat{G}_{BLS}^{-1}(\mathbf{q}) \qquad (5.12)$$

Comparing Eqs. (5.1) and (5.12), we get finally the direct susceptibility of the BLS:

$$\hat{d}_{BLS}(\mathbf{q},p) = \hat{\gamma}_{BLS}^{-1}(\mathbf{q},p) - \hat{G}_{BLS}^{-1}(\mathbf{q}) - \frac{p}{q^2}\hat{\Lambda}_{BLS}^{-1}(\mathbf{q},p). \qquad (5.13)$$

Inverting the matrices $\hat{\gamma}_{BLS}, \hat{G}_{BLS}, \hat{\Lambda}_{BLS}$ given by Eqs. (5.8), (5.9), (5.12), respectively (see Appendix 2), calculating $\hat{d}_{BLS}$ and identifying it with the direct susceptibility $\hat{d}$ of the polymeric system results in the final expression for the latter as the following diagonal block matrix:

$$\hat{d} = \begin{pmatrix} d_{\rho\rho} & d_{\rho\theta} & 0 \\ d_{\theta\rho} & d_{\theta\theta} & 0 \\ 0 & 0 & -\dfrac{p}{q^2\nu}\dfrac{\delta_{\alpha\beta} - n_\alpha n_\beta}{\Gamma_0} \end{pmatrix}. \qquad (5.14)$$

Here

$$d_{\rho\rho} = \frac{1}{\rho}\left(\frac{5}{2} + \frac{p}{q^2 D_0} - T^{-1}(\partial P/\partial \rho)_T - c_V\left(1 + \frac{p}{q^2 D_T}\frac{(\partial \theta/\partial \rho)_S^2}{T^2}\right) - pm\frac{\zeta}{T} - \frac{p^2}{q^2}\frac{m}{T}\right),$$

$$d_{\rho\theta} = d_{\theta\rho} = \frac{1}{\rho T}\left[c_V\left(1 + \frac{p}{q^2 D_T}\frac{(\partial \theta/\partial \rho)_S}{T}\right) - \frac{3}{2}\right], \quad d_{\theta\theta} = \frac{1}{\rho T^2}\left[\frac{3}{2} - c_V\left(1 + \frac{p}{q^2 D_T}\right)\right], \qquad (5.15)$$

where $\zeta = \nu + \mu = (\varphi + (4\eta/3))(m\rho)$.

It is worth to stress again that the submatrix $d^{\alpha\beta}$, which describes the "molecular flux field" appearing in non-equilibrium dense matter and acting on the broken links (monomer units), has the form characteristic for a low-molecular liquid and is inversely proportional to the viscosity $\upsilon$ of the BLS, which is quite natural because $\hat{d}$ describes the non-equilibrium properties of the system at small (segment) scales. The polymeric nature of the system will be incorporated into the properties of the corresponding submatrix of the structural susceptibility, $\gamma^{\alpha\beta}$. Moreover, $d^{\alpha\beta}$ turns out to be proportional to the well-known Oseen tensor [6]. Note that within the framework of DRPA the Oseen tensor naturally appears when describing the hydrodynamic interactions in the molecular field related to the viscous behavior of polymer melt. But modification of the density-density dynamic correlation function due to coupling between the density, temperature and flux fluctuations does not involve the Oseen tensor as we will see in the next section.

## 6 Sound dispersion and dynamic form-factor in compressible polymer melts

Now, when both the structural and direct susceptibility matrices are explicitly calculated and given by the expressions (4.15) and (5.14), (5.15), the inverse susceptibility of the monodisperse homopolymer melt is given by the basic DRPA relationship (3.7):

$$\hat{\alpha}^{-1}(\mathbf{q},p) = T(\hat{\gamma}^{-1}(\mathbf{q},p) - \hat{\mathbf{d}}(\mathbf{q},p))$$

$$= T \begin{pmatrix} \left(\frac{1}{\gamma_{\rho\rho}} + \frac{3}{2\rho} - \frac{p^2}{q^2\Gamma} - d_{\rho\rho}\right) & \left(-\frac{3}{2T\rho} - d_{\rho\theta}\right) & \frac{iq_\beta p}{q^2\Gamma} \\ \left(-\frac{3}{2T\rho} - d_{\theta\rho}\right) & \left(\frac{3}{2T^2\rho} - d_{\theta\theta}\right) & 0 \\ \frac{iq_\alpha p}{q^2\Gamma} & 0 & \frac{q_\alpha q_\beta}{q^2}\frac{1}{\Gamma} + \left(\delta_{\alpha\beta} - \frac{q_\alpha q_\beta}{q^2}\right)\left(\frac{1}{W} + \frac{p}{q^2\Gamma_0\nu}\right) \end{pmatrix} \quad (6.1)$$

and the desired block matrix of the generalized susceptibility

$$\hat{\alpha}(\mathbf{q},p) = \begin{pmatrix} \|\alpha^{(1)}_{ab}\| & \|\alpha^{(12)}_{a\beta}\| \\ \|\alpha^{(21)}_{\alpha b}\| & \|\alpha^{(2)}_{\alpha\beta}\| \end{pmatrix} \quad (6.2)$$

where the Latin and Greek indices run the values $(\rho,\theta)$ and $(x,y,z)$, respectively, could be readily found via inversion of the matrix (6.1) using the formulas (A2.2),(A2.3).

However, to analyze the dispersion of the longitudinal sound, which is the aim of our paper, it is sufficient to calculate the matrix $\boldsymbol{\alpha}^{(1)}$ only or, even more definitely, the component $\alpha_{\rho\rho}$ of the matrix. Indeed, let the external field $\vec{\varepsilon}_k(\mathbf{r},t)$ appearing in Eq. (2.1a) be of the form

$$\varepsilon_k^\rho(\mathbf{r},t) = A\delta(x)\exp(i\omega t), \quad \varepsilon_k^\theta(\mathbf{r},t) = 0, \quad \varepsilon_k^j(\mathbf{r},t)\varphi(\mathbf{r},t) = 0. \tag{6.3}$$

It follows from the definition (2.1) that the density change induced by such a field reads

$$\begin{aligned}\delta\rho(x,t) &= -A\int_{-\infty}^{t} dt' \int dz'dy'\,\alpha_{\rho\rho}(\mathbf{r}-\mathbf{r}',t-t')\exp(-i\omega t') \\ &= -A\int \alpha_{\rho\rho}(q,i\omega)\exp(i(qx-\omega t))dq/(2\pi) = iA\sum b_i \exp(i(q_i(\omega)x-\omega t)) \\ &= -A\sum b_i \exp(-\alpha x)\exp(i\omega((x/u_i(\omega))-t)),\end{aligned} \tag{6.4}$$

where $q_i$ is a pole of the function $\alpha_{\rho\rho}(q,i\omega)$ considered as a function of the complex variable $q$, $b_i = \mathrm{res}\,\alpha(q_i)$, and summation in Eq. (6.7) is over all poles of the function $\alpha_{\rho\rho}(q,i\omega)$. In other words, an oscillating in time with a frequency $\omega$ external field applied along a plane within a substance under consideration induces the harmonic compression-dilution waves, which are nothing but sound waves whose velocity $u$ and attenuation $\alpha$ are defined as

$$u_i^{-1}(\omega) = \mathrm{Re}(q_i(\omega)/\omega), \quad \alpha_i(\omega) = \mathrm{Im}\,q_i(\omega) \tag{6.5}$$

and fully determined by the analytical behavior of the function $\alpha_{\rho\rho}(q,i\omega)$.

As consistent with the rules of the block matrix inversion (see Appendix 2), the matrix expression for $\mathbf{\alpha}^{(1)}$ reads:

$$\mathbf{\alpha}^{(1)} = \begin{pmatrix} \alpha_{\rho\rho} & \alpha_{\rho\theta} \\ \alpha_{\theta\rho} & \alpha_{\theta\theta} \end{pmatrix} = \begin{pmatrix} \left(\gamma_{\rho\rho}^{-1}(q,p)-\tilde{d}_{\rho\rho}\right) & -\tilde{d}_{\rho\theta} \\ -\tilde{d}_{\theta\rho} & -\tilde{d}_{\theta\theta} \end{pmatrix}^{-1}, \tag{6.6}$$

where

$$\tilde{d}_{\rho\rho} = \frac{1}{\rho}\left(1 - T^{-1}(\partial P/\partial\rho)_T + \frac{p}{q^2 D_0} - c_V\left(1 + \frac{p}{q^2 D_T}\frac{(\partial\theta/\partial\rho)_S^2}{T^2}\right) - pm\frac{\zeta}{T} - \frac{p^2}{q^2}\frac{m}{T}\right),$$

$$\tilde{d}_{\rho\theta} = \tilde{d}_{\theta\rho} = \frac{c_V}{\rho T}\left(1 + \frac{p}{q^2 D_T}\frac{(\partial\theta/\partial\rho)_S}{T}\right), \quad \tilde{d}_{\theta\theta} = -\frac{c_V}{\rho T^2}\left(1 + \frac{p}{q^2 D_T}\right). \tag{6.7}$$

It follows from (6.6) that, in particular,

$$\alpha_{\rho\rho} = \left(\gamma_{\rho\rho}^{-1}(q,p) - \tilde{d}(q,p)\right)^{-1}, \tag{6.8}$$

where

$$\tilde{d}(q,p) = \tilde{d}_{\rho\rho} - \left(\tilde{d}_{\rho\theta}^2/\tilde{d}_{\theta\theta}\right) = \frac{1}{\rho}\left(1 + \frac{p}{q^2 D_0}\right) - \frac{1}{\rho u_0^2}\left(c_T^2 + \frac{p^2}{q^2} + p\frac{c_S^2 - c_T^2}{q^2 D_T + p} + p\zeta\right), \tag{6.9}$$

where $c_T = \left(m^{-1}(\partial P/\partial \rho)_T\right)^{1/2}$, $c_S = \left(m^{-1}(\partial P/\partial \rho)_S\right)^{1/2}$ are the isothermal and adiabatic sound velocities for the BLS and $u_0 = (T/m)^{1/2}$ is a characteristic heat velocity of the BLS particles. It follows from Eqs. (6.8), (6.9) that the poles $q_i$ of the function $\alpha_{\rho\rho}(q, i\omega)$ are zeros of the dispersion equation

$$u_0^2 V(q,p) + \left(c_S^2 - \frac{\omega^2}{q^2} + i\omega\zeta - \frac{(c_S^2 - c_T^2)q^2 D_T}{q^2 D_T + i\omega}\right) = 0, \qquad (6.10)$$

$$V(q,\omega) = \frac{\rho}{\gamma_{\rho\rho}(q,i\omega)} - \frac{\rho}{\gamma_{BLS}^{\rho\rho}} = \frac{\rho}{\gamma_{\rho\rho}(q,i\omega)} - \left(1 + \frac{i\omega}{q^2 D_0}\right). \qquad (6.11)$$

Eqs. (6.8), (6.10) are rather important and deserve some discussion. First, the scalar equation (6.8) has the same structure as the basic DRPA matrix Eq. (3.7): the structural polymer and energetic interaction contributions into the inverse collective susceptibility are additive and described by the terms $\gamma_{\rho\rho}^{-1}(q,p)$ and $-\tilde{d}(q,p)$, respectively. It is in this form that the dynamic RPA was introduced in the work [19] by one of us, even though the approximation used for actual calculations in this work was

$$\tilde{d}(q,p) \approx d(q,0) = \rho^{-1}\left(1 - T^{-1}(\partial P/\partial \rho)_T\right). \qquad (6.9a)$$

Second, it could appear somewhat strange that the quantity $\tilde{d}(q,p)$ characterizing both static and dynamic interactions between the particles of the BLS contains a term depending on the self-diffusion coefficient $D_0$ of these particles. Moreover, the relative magnitude of this term is enormous for $\omega \ll u_0^2/D_0 \sim 10^{15}$ s$^{-1}$ i.e. for any reasonable frequency. But, in fact, $D_0$ also depends on the interaction between the particles and their density (if there is no interaction the particles move freely, do not scatter and, therefore, do not reveal any Brownian motion). Besides, $D_0$ also appears in the structural susceptibility $\gamma_{\rho\rho}(q,i\omega)$. For the BLS the difference of these two $D_0$-dependent terms making the function $V(q,p)$ is strictly zero by definition and the dispersion equation takes the form

$$c_T^2 - \frac{\omega^2}{q^2} + i\omega\left(\frac{c_S^2 - c_T^2}{q^2 D_T + i\omega} + \zeta\right) = 0, \qquad (6.10a)$$

which is well known in the theory of simple liquids [42].

On the contrary, for polymer melts the function $V(q,p)$ differs from zero and it is this fact that results into the sound behavior peculiar for the latter. Indeed, let us calculate the quantity $V(q,p)$ using Eqs. (4.11):

$$V(Q,K) = \frac{1}{N}\left( \frac{1}{f_D(2Q)} + K\left(1 + \coth Q - \frac{1}{Q}\right) + \frac{KF(Q,K)}{Q} \right) - 1, \quad (6.12)$$

where the reduced variables $Q$ and $K$ are defined above. Taking into account that the length of any sound waves is very large as compared to the size $R$ of a polymer coil and, therefore, the value of $Q \sim (qR)^2$ is rather small, we can expand the function (6.11) in powers of $Q$ and keep the first non-vanishing terms of these expansions. It results in the following rather accurate approximation:

$$V(Q,K) \approx (N^{-1} - 1) + N^{-1}(K + Q\, g(K)), \quad (6.12)$$

where we introduced the definition

$$g(K) = 1 + \frac{K}{3} - \frac{(\sqrt{K}\coth\sqrt{K} - 1)}{K} \approx \begin{cases} K/3, & K \to \infty \\ 2/3, & K \to 0 \end{cases}. \quad (6.13)$$

and took into account that for the Rouse chains the equality $ND_N = D_0$ holds. Thus, the dispersion equation for homopolymer monodisperse melts reads

$$\tilde{c}_S^2 - \frac{\omega^2}{q^2} + i\omega(\zeta + \zeta_1) + q^2\left( u_0^2 \frac{b^2}{12} g\!\left(\sqrt{i\omega\tau_R/2}\right) - D_T\!\left(\frac{c_S^2 - c_T^2}{q^2 D_T + i\omega}\right) \right) = 0 \quad (6.10b)$$

where $\zeta_1 = u_0^2 \tau_R/(2N) = Nb^2 u_0^2/(12 D_0)$ is a specific polymer contribution into the volume viscosity and $\tilde{c}_S^2 = c_S^2 - u_0^2(1 - N^{-1})$ is a polymer characteristic sound velocity. It is worth to notice that $\zeta_1$ is nothing but a remnant of the $q$-dependence of the high frequency limit $\lambda_\infty(Q)$ of the polymer kinetic coefficient, or, in other words, of the non-local nature of the dissipative processes in polymers.

Now, before to continue our analysis, it is worth to remind here the typical (very rough) numerical values of the parameters that appear in Eq. (6.15): $u_0 < c_T < c_S \sim 10^3$ m/s, $\zeta \sim D_T \sim 10^{-7}$ m$^2$/s, $D_0 \sim 10^{-9}$ m$^2$/s, $b \sim 10^{-9}$ m, and, thus, $\zeta_1 \sim N\, 10^{-4}$ m$^2$/s. Thus, the inequality $\zeta \ll \zeta_1$ holds, which could be considered as a particular case of the following general assertion (to be verified later): the dissipation due to the BLS effects (thermoconductivity $D_T$ and volume viscosity $\zeta$) is negligible as compared to the specifically polymer dissipation related to the internal (Rouse) modes of the polymer chains and all the BLS terms could be omitted.

The final dispersion equation in this approximation reads

$$\tilde{c}_S^2 - \frac{\omega^2}{q^2} + i\omega\zeta_1 + q^2 u_0^2 \frac{b^2}{12} g\!\left(\sqrt{i\omega\tau_R/2}\right) = 0 \quad (6.14)$$

To describe the roots of the dispersion equation (6.14) it is convenient to use the complex sound velocity $c_1 = \omega/q$ rather than the complex wave vector $q$ (the frequency $\omega$ is a real quantity). These roots are determined by the expression

$$c_1^2 = \left( \tilde{c}_S^2 + i\omega\zeta_1 \pm \sqrt{\left(\tilde{c}_S^2 + i\omega\zeta_1\right)^2 + (b\omega u_0)^2 \, g\left(\sqrt{i\omega\tau_R}\right)/3} \right)/2. \tag{6.15}$$

The positive sign of the square root in (6.15) corresponds to a basically real velocity. It describes the sound waves with a small dissipation. On the contrary, the negative sign does to a basically imaginary velocity and it describes the intrinsically dissipative propagation of heat. Being interested in this paper by the sound propagation only, we restrict ourselves in this paper by the first root, Therewith, it is easy to see that the term proportional to the function $g$ could be neglected if the condition $\omega\tau_R \ll (\zeta_1/(bu_0))^2 \sim 10^4 N^2$ holds, which is the case for any reasonable frequencies. Thus, the complex sound velocity simply reads

$$c_1 = \tilde{c}_S \left(1 + \omega^2\tau^2\right)^{1/4} \exp(i\Omega/2), \qquad \Omega = \arctan(\omega\tau), \tag{6.16}$$

where the characteristic time

$$\tau = \zeta_1/\tilde{c}_S^2 = N b^2 u_0^2 / \left(12 D_0 \tilde{c}_S^2\right) \tag{6.17}$$

is a sort of geometric mean ($\tau_R/\tau \sim \tau/\tau_1 \sim N$) of the Rouse time and a "monomeric" time $\tau_1 = b^2/(6 D_0)$.

The observable sound velocity $u$ and the absorption coefficient $\alpha/\omega^2$ are related to the complex sound velocity $c_1$ as

$$u(\omega) \approx \left(\text{Re}(1/c_1)\right)^{-1} = \tilde{c}_S \, 2^{1/2} \left(1 + \omega^2\tau^2\right)^{1/2} \Big/ \left(1 + \left(1 + \omega^2\tau^2\right)^{1/2}\right)^{1/2},$$

$$\Gamma = \frac{\alpha}{\omega^2} = \text{Im}\frac{1}{\omega c_1} = \frac{1}{\tilde{c}_S} \frac{\tau}{\left(1 + \omega^2\tau^2\right)^{1/2} \left(1 + \left(1 + \omega^2\tau^2\right)^{1/2}\right)^{1/2}}, \tag{6.18}$$

It follows from Eq. (6.18) that $u$ and $\alpha/f^2$ approach constant values in the limit $\omega \to 0$:

$$u(0) = \tilde{c}_S, \qquad \left(\frac{\alpha(\omega)}{\omega^2}\right)\bigg|_{\omega=0} = \frac{\tau}{2^{1/2} \tilde{c}_S} = \frac{N b^2 u_0^2}{12\sqrt{2} D_0 \tilde{c}_S^3}. \tag{6.19}$$

Thus, the numerical value of the low-frequency sound velocity in a polymer melt is of the same order of magnitude as that in the corresponding low-molecular melt (BLS): $u(0) = \tilde{c}_S \sim c_S$. At the same time, the low-frequency absorption coefficient given by Eq. (6.18) is proportional to the degree of polymerization $N$ and much greater than the corresponding coefficient for the BLS $\Gamma(\omega = 0) = \left(\zeta + D_T\left(1 - (c_T/c_S)^2\right)\right)/\left(2^{1/2} c_S^3\right)$.

The reduced (normalized by their zero-frequency values) sound velocity and absorption

coefficient

$$\tilde{u}(\omega) = \frac{u(\omega)}{u(0)} = \left(\frac{2(1+\omega^2\tau^2)}{1+(1+\omega^2\tau^2)^{1/2}}\right)^{1/2}, \quad \tilde{\Gamma} = \frac{\alpha(\omega)/\omega^2}{(\alpha(\omega)/\omega^2)\big|_{\omega=0}} = \left(\frac{2}{\left((1+\omega^2\tau^2)+(1+\omega^2\tau^2)^{3/2}\right)}\right)^{1/2} \quad (6.18a)$$

are plotted against the reduced frequency $\omega^* = \omega\tau$ in Fig. 1.

It is worth to notice here in the high-frequency region $\omega \geq N^2/\tau_R \sim \tau_1$ corresponding to the Rouse modes of the highest order, the relaxation processes related to the individual segments of polymer chains are expected to cause considerable dispersion of the direct susceptibility (see section 5), which is not taken into account in the present work. Besides, the expression (4.10) for the structural susceptibility, which corresponds to neglecting the discrete nature of $N$-mers, also becomes inapplicable in this region. Thus, the expressions for the sound velocity and absorption coefficient given by Eqs. (6.18) make sense in the frequency range $\omega\tau_1 \ll 1$ or $f \ll 10^9\text{-}10^{10}$ Hz.

To finish with calculations, we consider now the correlation function of the density-density fluctuations in the co-ordinate and Fourier representations

$$S_{\rho\rho}(\mathbf{r},t) = \langle \delta\rho(\mathbf{r},t)\delta\rho(0,0)\rangle, \quad S_{\rho\rho}(q,\omega) = \int_{-\infty}^{\infty} dt \int d\mathbf{r} \exp(i\mathbf{q}\mathbf{r} - i\omega t) S_{\rho\rho}(\mathbf{r},t) \quad (6.20)$$

It follows from the definitions (2.4), (6.8) and (6.20) that in the approximation we used to derive the dispersion equation (6.14) the dynamic form-factor reads

$$S_{\rho\rho}(q,\omega) = -2\,\text{Im}\,\alpha(q,i\omega)/\omega = 2\rho u_0^2 \zeta_1 / \left[\left(\tilde{c}_S^2 - (\omega^2/q^2)\right)^2 + \omega^2\zeta_1^2\right]. \quad (6.21)$$

If the value of the wave number $q$ is small enough ($q < q_{\lim} = \sqrt{2}\,\tilde{c}_S/\zeta_1 \sim 10^7/N$ m$^{-1}$) then the function (6.21) has two symmetric maximums corresponding to the Brillouin doublet. It is worth to notice that the velocity experimentally determined from this maximum location

$$\bar{c} = \omega_{\max}/q \quad \bar{c} = \sqrt{\tilde{c}_S^2 - (\zeta_1^2 q^2/2)} \quad (6.22)$$

could be, due to anomalously large volume viscosity $\zeta_1$ of polymer melts, considerably less than the zero-frequency sound velocity $\tilde{c}_S$. This Brillouin maximum shifts to zero with increase of the degree of polymerization $N$ and could disappear at all if the scattering angle is not small $(q > q_{\lim})$. Until then the function (6.21) does not reveal any central Landau-Placzek maximum because the latter is due to the term $\sim q^2 D_T/(q^2 D_T + i\omega)$ irrelevant for the Brillouin scattering and, therefore, skipped in (6.21).

The precise expression for the Fourier inversion of the dynamic form-factor (6.21) (i.e. the correlation function $S_{\rho\rho}(\mathbf{r},t)$) is rather cumbersome. However, it has two simple asymptotics:

$$S_{\rho\rho}(\mathbf{r},t) = \begin{cases} A\delta(\mathbf{r}), & t=0 \\ A\dfrac{(\tau/(2\pi t))^{3/2}}{(\tilde{c}_S \tau)^3} \exp\left(-\dfrac{t}{2\tau}\left(1-\dfrac{x}{\tilde{c}_S t}\right)^2\right), & t \gg \tau \end{cases} \qquad (6.23)$$

where $A = \rho u_0^2 / (c_T^2 - u_0^2(1-N^{-1}))$. As is seen clearly from (6.23), the fluctuation relaxation in polymer melts proceeds via propagation of the density waves, the latter being decayed exponentially with time if their velocity $x/t$ differs from the sound velocity $\tilde{c}_S$ and power-like (as $t^{-3/2}$) if their velocity equals the zero-frequency sound velocity $\tilde{c}_S$.

Therewith, the time $\tau$ that characterizes the rate of the fluctuation relaxation in (6.23) as well as the scale of the sound dispersion in (6.18) is defined by (6.17). It is worth to stress again that this time is $N$ times less than the time $\tau_R \sim N^2 b^2/D_0$ characterizing the Rouse dynamics of single $N$-mers forming the melt under consideration. It is natural to refer to this important fact as "dynamic screening" in polymer melts because of its similarity to the well-known "static screening" i.e. decrease of the correlation radius from the value $R \sim Na$ characterizing the scale of a single $N$-mer to $R \sim r_0$ for polymer melts ($r_0$ is the correlation radius in the BLS). The dynamic screening, which, up to our knowledge, for polymer melts is first rigorously proved in the present work, is caused by the presence of strong interchain interaction in dense polymer systems, as is the static one. Its importance follows from the fact that, when basing on the existing single-chain dynamics theories [1]-[4], one would have to interpret the characteristic time extracted from any frequency-dependent dynamic data for polymer melts as the Rouse one.

Now, let us compare the theoretical results we have obtained in this section and the experimental data. The broadband measurements of the sound velocity $u(\omega)$ are rather scarce because of the high demands on accuracy, as mentioned in the very recent review on the ultrasonic spectrometry of liquids [44]. Nevertheless, some measured and estimated data found in Ref. [45] as well as indirect evidences stemming from the dynamic shear measurements reported in Ref. [7] clearly show that $u$ increases with $\omega$ in the megahertz range in agreement with our result presented in Eq. (6.18). Next, as is seen from Fig.1, the ω-dependence of the absorption coefficient in log-log co-ordinates becomes steeper when the frequency increases, which is in a qualitative agreement both with the experiments reported in Ref. [2], [5] and preceding theories [3], [4].

However, the high-frequency asymptotic dependence $\Gamma \sim \omega^{-3/2}$, which follows from our theory as consistent with Eq. (6.18), is considerably stronger than that observed experimentally. In particular, the sound absorption coefficient for the poly(phenyl methyl siloxane) melt of disentangled chains ($\overline{M} = 5000$, $N = 28$) measured in the range 0.5-100 Mhz was well described by the dependence $\Gamma \sim \omega^{-0.3}$ [5]. This discrepancy cannot be attributed to the fact that the experimental data correspond to the region $\omega\tau \sim 1$ because, first, the experimentally measured relative frequency increase of the sound velocity is much less than that of the absorption coefficient, whereas they should be of the same order of magnitude for $\omega\tau \leq 1$, and, second, the observed change of $\Gamma$ in this interval is two order of magnitude [5].

To understand the sources of such a discrepancy between our theory and experiment let us remember that the physical origin of the anomalously large volume viscosity $\zeta_1$ in polymer melts is a subtle dismatch between the inverse kinetic coefficients of the ideal BLS and Rouse chains systems multiplied by the great factor $K/Q$ (see discussion on the pages 22-23). In turn, this dismatch is due to non-local character of the single chain dynamics resulting into a residual $Q$-dependence (spatial dispersion) of the high-frequency limit of the inverse kinetic coefficient of $N$-mers. Therefore, one could expect the dynamic properties of polymer melts to be rather sensitive to the properly averaged particular dynamic properties of the chains forming the melt. In particular, as shown by one of us and Semenov [20], the dynamic form-factor of polydisperse melt reveals considerable frequency dispersion and hence its evolution is rather different from that in monodisperse melt. As applied to the problem under consideration, it could lead to emergence of two different characteristic times appearing in the frequency dependences of the sound absorption coefficient and velocity, these times being involved different averaging of $N$-dependent characteristics and, thus, having different orders of magnitude. Thus, one of the plausible reasons of the discrepancy could be that the experimental samples are polydisperse (the polydispersity index of 1.5 was reported in Ref. [5]) whereas the results presented in this paper are related to monodisperse polymer melt. We suppose to study sound propagation in polydisperse melts elsewhere.

One more interesting option is that the single chain dynamics could differ from the Rouse one. For instance, it could include some cooperative gauche-trans transitions of monomer units [46] or even soliton-like excitations (see review [47] and reference therein), which also could modify the sound frequency dispersion substantially as compared to the Rouse model.

## 7 Conclusion

Let us summarize the gist of the approach presented in this paper and the problems solvable and appearing within the framework of our approach.

The new way of a unified description of all dissipative processes in polymer systems we have presented here is based on three simple and physically clear ideas: *i*) the global response and dynamic properties of individual macromolecules forming a polymer system are basically the same as for the single ones. *ii*) the actual difference between the dynamical properties of the concentrated and diluted polymers systems is that the fields inducing the single-chain relaxation are strongly renormalized dynamically as compared to the original external fields (because of a strong interchain interaction) and depend on how much is the whole system deviated from its equilibrium state. *iii*) the coupling between the renormalized fields and the magnitude of local deviations from equilibrium does not depend on the polymer structure of the system.

These ideas in common are nothing but a straightforward generalization of the famous Flory concept that polymer chains in melts are Gaussian, which has transformed currently into the general framework of the self-consistent field description of polymer systems. The new and key step in this direction made in this work is to describe the coupling between the renormalized fields and local deviations from equilibrium via a new $(5 \times L) \times (5 \times L)$ direct susceptibility matrix $\hat{\mathbf{d}}$ to be explicitly calculated from *L*-fluid hydrodynamics. (Up to our knowledge, the matrix $\hat{\mathbf{d}}$, which naturally appears in the polymer theory as shown above, never has been introduced yet, even though one of its $(L \times L)$ sub-blocks, $\hat{d}$, is for a long time known in the theory of simple liquids as the matrix of direct correlation functions.)

Implementation of these ideas is Eq. (3.7) relating the observable $(5 \times L) \times (5 \times L)$ matrix of the collective susceptibilities $\hat{\boldsymbol{\alpha}}$ to the model-dependent structural susceptibility $\hat{\boldsymbol{\gamma}}$ and direct susceptibility $\hat{\mathbf{d}}$. All the accompanying calculations are straightforward but numerous and sometimes cumbersome due to the high-rank matrix nature of the relevant equations. This is why our main objective here was just to show that the proposed procedure does work and leads to some new and reasonable results. in the simplest situation of the density relaxation and fluctuations. More precisely, to demonstrate the concept of the direct susceptibility matrix $\hat{\mathbf{d}}$ in action we have calculated it for *1*-fluid hydrodynamics. As far as the structural susceptibility $\hat{\gamma}$ is concerned, we restricted ourselves to the case when the chains are short enough to obey the Rouse rather than the reptation dynamics. Then, to calculate the proper components of the dynamic structural susceptibility of Rouse chains, we have shown (Appendix 1) that the dynamic correlation function $\langle \exp(i\mathbf{q}[\mathbf{r}(s,t) - \mathbf{r}(s',0)]) \rangle$ for the Rouse model obeys an equation of the same type as that for the reptation model. This somewhat unexpected fact has lead to a comparatively simple expression (4.10), which may be considered as an alternative to the well-known representation of the dynamic structural factor $S_{\rho\rho}(q,t)$ in the form of asymptotic series [6]. Finally, we used the obtained expressions (4.14) for $\hat{\gamma}$ and (5.14) for $\hat{\mathbf{d}}$ to calculate explicitly the sound velocity and absorption coefficient frequency dispersion for monodisperse Rouse homopolymer melt. Besides, we have calculated the dynamic form-factor for the melt and analyzed some peculiarities of the Brillouin scattering in it.

The most appealing feature of the new approach is that it solves, in principle, an old classic problem, which is as follows. How does a chain inside of a bulk amorphous or liquid sample know that there is a shearing perturbation on the sample surface? The answer is that this perturbation generates some dynamically renormalized fields propagating into the sample interior and eventually affecting the single chain conformations in the whole sample.

We suppose to address this problem as well as a variety of other problems solvable within the presented approach (and, first of all, to study contribution of the Brillouin scattering on the sound waves under microphase separation in block copolymers) elsewhere.

The authors acknowledge useful and stimulating discussions with V. Shilov and M. Gotlieb and financial support by INTAS (project No 99-01852).

**Appendix 1. The dynamic structural factor in the Rouse model**

It is generally accepted [6] that the dynamic behavior of a monodisperse melt with the degree of polymerization less than the entanglement threshold is well described by the Rouse model. In this bead-spring model, it is assumed that each segment of a polymer chain (which may include several monomer units depending on the stiffness of a modeled macromolecule) moves under the action of the elastic force from segments neighboring in the chain, the effective friction force, and the random force from segments neighboring in space. Omitting the inertial term insufficient at the space-time scale characterizing collective segment motions, one may write for $i$-th internal segment of a given chain

$$\xi \frac{d\mathbf{r}_i(t)}{dt} = \frac{3T}{b^2}[\mathbf{r}_{i+1}(t) - 2\mathbf{r}_i(t) + \mathbf{r}_{i-1}(t)] + \mathbf{f}_i(t). \tag{A1.1}$$

where $\mathbf{r}_i(t)$ denotes the position of $i$-th segment at time $t$, $\xi$ is the friction constant of the melt, $b$ is the segment size, $\mathbf{f}_i(t)$ is the random force acting on $i$-th segment and obeying the Gaussian distribution: $\langle \mathbf{f}_i(t) \rangle = 0$, $\langle f_i^\alpha(t) f_j^\beta(t') \rangle = 2\xi T \delta_{\alpha\beta} \delta_{ij} \delta(t-t')$.

If the total number of segments in the chain $N \gg 1$, it is convenient to replace $i$ with a continuous variable $s = ib$, $0 \leq s \leq Nb$. In this case Eq. (A1.1) takes the form

$$\frac{\partial \mathbf{r}(s,t)}{\partial t} = 3D_0 \frac{\partial^2 \mathbf{r}(s,t)}{\partial s^2} + \frac{\mathbf{f}(s,t)}{\xi}, \tag{A1.2}$$

where $D_0 = T/\xi$ has the meaning of the segment diffusivity in the BLS.

Let us define the molecular dynamic correlator

$$g(\mathbf{q},s,s',t) = \langle \exp(i\mathbf{q}[\mathbf{r}(s,t) - \mathbf{r}(s',0)]) \rangle, \tag{A1.3}$$

which is related to the dynamical structural function (4.9) as follows:

$$\sigma_{\rho\rho}(\mathbf{q},t) = (\nu/b^2) \int_0^{Nb} ds \int_0^{Nb} ds' \, g(\mathbf{q},s,s',t) \tag{A1.4}$$

where $\nu$ is the number of chains per unit volume.

Differentiating (A1.3) with respect to time, we get

$$\frac{\partial g(\mathbf{q},s,s',t)}{\partial t} = \left\langle i\mathbf{q} \frac{\partial \mathbf{r}(s,t)}{\partial t} \exp(i\mathbf{q}[\mathbf{r}(s,t) - \mathbf{r}(s',0)]) \right\rangle \tag{A1.5}$$

On the other hand, the double differentiation with respect to $s$ yields

$$\frac{\partial^2 g(\mathbf{q},s,s',t)}{\partial s^2} = \left\langle \left( i\mathbf{q} \frac{\partial^2 \mathbf{r}(s,t)}{\partial s^2} - \left( \mathbf{q} \frac{\partial \mathbf{r}(s,t)}{\partial s} \right)^2 \right) \exp(i\mathbf{q}[\mathbf{r}(s,t) - \mathbf{r}(s',0)]) \right\rangle \tag{A1.6}$$

For $qb \ll 1$, i.e. at distances much greater than the segment size $b$, the term in (A1.6) proportional to $q^2$ is to be neglected since $(\mathbf{q}\partial\mathbf{r}(s,t)/\partial s)^2 \cong q^2 \ll (\mathbf{q}\partial^2\mathbf{r}(s,t)/\partial s^2) \cong q/b$.

Using Eq. (A1.2) we obtain from Eq. (A1.5), (A1.6)

$$\frac{\partial g(\mathbf{q},s,s',t)}{\partial t} = 3D_0 \frac{\partial^2 g(\mathbf{q},s,s',t)}{\partial s^2} + \left\langle i\mathbf{q}\frac{\mathbf{f}(s,t)}{\xi}\exp(i\mathbf{q}[\mathbf{r}(s,t)-\mathbf{r}(s',0)])\right\rangle. \quad (A1.7)$$

Expanding the exponent and making use of the properties of the random Gaussian variables $\mathbf{f}(s,t)$ and $\mathbf{r}(s,t)-\mathbf{r}(s',0)$, we may represent Eq. (A1.7) in the form

$$\frac{\partial g(\mathbf{q},s,s',t)}{\partial t} = 3D_0 \frac{\partial^2 g(\mathbf{q},s,s',t)}{\partial s^2} - \frac{q^2}{3\xi}\langle\mathbf{f}(s,t)\mathbf{r}(s,t)\rangle g(\mathbf{q},s,s',t). \quad (A1.8)$$

Using the solution of Eq. (A1.2) in terms of the normal coordinates (Rouse modes) [6] it is easy to calculate the correlator

$$\langle\mathbf{f}(s,t)\mathbf{r}(s',t)\rangle = \frac{3T}{N}\left(1 + 4\sum_{m=1}^{\infty}\cos\left(\frac{\pi s m}{N}\right)\cos\left(\frac{\pi s' m}{N}\right)\right) \approx 6T\delta(s-s'), \quad (A1.9)$$

where $m$ numerates the modes.

Hence the correlator $\langle\mathbf{f}(s,t)\mathbf{r}(s,t)\rangle$ entering (A1.8) diverges. Clearly, this is an artifact arising when ordinal numbers of segments are replaced by the continuous coordinate along the chain, $s$ (see Eqs. (A1.1), (A1.2)). Indeed, for a polymer chain consisting of $N$ segments (beads) connected by chemical links (springs), a motion of any segment can be resolved into $N$ independent modes, each mode being correlated with a corresponding random force. There is no physical reason neither for any of these $N$ correlators nor for their sum to be infinite.

To calculate the value of $\langle\mathbf{f}(s,t)\mathbf{r}(s,t)\rangle$ explicitly, one should solve the discrete equations of motion (A1.1) rather than its continuous analogue Eq. (A1.2). In this work, we want only to estimate this correlator. To this aim, we cut the summation in (A1.9) at $m = N$ thus finding

$$\langle\mathbf{f}(s,t)\mathbf{r}(s,t)\rangle \cong \frac{3T}{N}\left(1 + 4\sum_{m=1}^{N}\cos^2\left(\frac{\pi s m}{N}\right)\right) \leq 6T. \quad (A1.10)$$

Now we estimate the first and second terms on the right-hand side of Eq. (A1.8) to be of the order $D_0(\mathbf{q}\partial^2\mathbf{r}/\partial s^2)g \cong D_0 g q/b$ (see Eq. (A1.6)) and $q^2(T/\xi)g = q^2 D_0 g$, respectively. Therefore, at $qb \ll 1$ we can neglect the latter obtaining

$$\frac{\partial g(\mathbf{q},s,s',t)}{\partial t} = 3D_0 \frac{\partial^2 g(\mathbf{q},s,s',t)}{\partial s^2}. \tag{A1.11}$$

The initial condition for Eq. (A1.11) reflects the fact that in the Rouse model the equilibrium distance between any two points of a chain is a random Gaussian variable [6]:

$$g(\mathbf{q},s,s',0) = \exp(-q^2 b|s-s'|/6). \tag{A1.12}$$

The boundary conditions may be formulated if notice that the motion of chain ends is uncorrelated with the conformation of the rest chain. This property has been exploited for deriving the boundary conditions for the correlator $g$ of a reptating chain. It is applicable in the Rouse model as well, since disentangled chains also take random conformations in space. Following Ref. [6], we may write

$$\frac{\partial}{\partial s}g(\mathbf{q},s=L,s',t) = -\frac{1}{6}q^2 b g(\mathbf{q},s=L,s',t),$$
$$\frac{\partial}{\partial s}g(\mathbf{q},s=0,s',t) = \frac{1}{6}q^2 b g(\mathbf{q},s=0,s',t). \tag{A1.13}$$

It is worth pointing out that not only the boundary conditions (A1.13) but the equation (A1.11) for $g$ itself as well as the initial condition (A1.12) coincide with those for the reptation model. The only difference consists in the definition of the diffusion coefficient entering Eq. (A1.11).

The solution of Eq. (A1.11) - (A1.13) is well-known [6]. Applying the Laplace transformation, one gets [19]

$$g(\mathbf{q},s,s',p) = \int_0^\infty dt\, g(\mathbf{q},s,s',t)\exp(-pt) = \frac{\tau_R}{2}\frac{Q}{Q^2-K}\left(\frac{\exp(-\sqrt{K}|\theta-\theta'|)}{\sqrt{K}} - \frac{\exp(-Q|\theta-\theta'|)}{Q}\right)$$
$$-\frac{Q\exp(-\sqrt{K})}{\sqrt{K}(Q+\sqrt{K})}\left(\frac{\cosh(\sqrt{K}\theta)\cosh(\sqrt{K}\theta')}{\sqrt{K}\sinh\sqrt{K}+Q\cosh\sqrt{K}} + \frac{\sinh(\sqrt{K}\theta)\sinh(\sqrt{K}\theta')}{\sqrt{z}\cosh\sqrt{z}+Q\sinh\sqrt{K}}\right), \tag{A1.14}$$

where $\theta = (2s-Nb)/(Nb)$, $\theta' = (2s'-Nb)/(Nb)$, $Q = q^2 b^2 N/12$, $K = pN^2 b^2/(12D_0)$.

Substituting (A1.14) into (A1.4) and taking into account that $\nu = \rho/N$, $\rho$ being the number density of segments, we get the dynamic structural function:

$$\sigma_{\rho\rho}(\mathbf{q},p) = \frac{\rho N \tau_R}{2K}\left(\left(\frac{1}{Q} - \frac{1}{Q^2(1+\coth Q)}\right) - \frac{Q^2}{Q^2-K}\left(\frac{1}{K+Q\sqrt{K}\coth\sqrt{K}} - \frac{1}{Q^2(1+\coth Q)}\right)\right). \tag{A1.15}$$

The behavior of the structural function $\sigma_{\rho\rho}(\mathbf{q},t)$ at $t \to \infty$ is determined by the expansion of $\sigma_{\rho\rho}(\mathbf{q},p)$ near its pole with the maximal real part $K \approx -Q$. It is easy to find

$$\sigma_{\rho\rho}(\mathbf{q},t) \cong \rho N \exp\left(-\frac{2Qt}{\tau_R}\right) = \rho N \exp(-q^2 Dt), \quad D = \frac{D_0}{N}. \tag{A1.16}$$

The structural function at small times ($t \ll \tau_R$) is obtained by expanding $\sigma_{\rho\rho}(\mathbf{q},p)$ at $K \to \infty$. At $Q^2 \ll K$ we have

$$\sigma_{\rho\rho}(\mathbf{q},p) = \frac{\rho N \tau_R}{2}\left(\frac{1}{K}\left(\frac{1}{Q} - \frac{1}{Q^2(1+\coth Q)}\right) - \frac{1}{K^2(1+\coth Q)} + \frac{Q^2}{K^3(1+\tanh Q)} + O(K^{-4})\right). \tag{A1.17}$$

It determines the behavior of $\sigma_{\rho\rho}(\mathbf{q},t)$ at $Q \gg 1$ ($q^2 R^2 \gg 1$):

$$\sigma_{\rho\rho}(\mathbf{q},t) = \rho N\left(\frac{1}{Q} - \frac{t}{\tau_R} + \frac{Q^2 t^2}{\tau_R^2} + O(t^3)\right) \tag{A1.18}$$

and at $Q \ll 1$ ($q^2 R^2 \ll 1$):

$$\sigma_{\rho\rho}(\mathbf{q},t) = \frac{\rho N \tau_R}{2}\left(1 - \frac{2Qt}{\tau_R} + \frac{2Q^2 t^2}{\tau_R^2} + O(t^3)\right). \tag{A1.19}$$

If $K \to \infty$ but $Q^2 \gg K$, then we get from Eq. (A1.15)

$$\sigma_{\rho\rho}(\mathbf{q},p) \approx \frac{\rho N \tau_R}{2}\left(\frac{1}{KQ} - \frac{1}{K^{3/2}Q}\right) \text{ that corresponds to } \sigma_{\rho\rho}(\mathbf{q},t) \approx \rho N\left(1 - 2Q\sqrt{\frac{2t}{\pi \tau_R}}\right).$$

These expressions are in agreement with the asymptotics of the dynamic structural function calculated via solving the Langevin equations (A1.1) for the radius-vectors of segments [6]. As to our knowledge, the way of calculating $\sigma_{\rho\rho}(\mathbf{q},p)$ presented in this work has not been implemented so far.

**Appendix 2. Inversion of the block matrices.**

Let $\mathbf{A}$ be a block $(m+n)\times(m+n)$ matrix consisting of two diagonal square blocks (($m\times m$) $\mathbf{G}_1$ and ($n\times n$) $\mathbf{G}_2$) and two off-diagonal (generally, rectangular) blocks (($m\times n$) $\mathbf{\Gamma}_{12}$ and ($n\times m$) $\mathbf{\Gamma}_{21}$). The matrix $\mathbf{A}^{-1}$ inverse to the block matrix $\mathbf{A}$ has, evidently, the same block structure:

$$\mathbf{A} = \begin{pmatrix} (\mathbf{G}_1)_{ij} & -(\mathbf{\Gamma}_{12})_{i\beta} \\ -(\mathbf{\Gamma}_{21})_{\alpha j} & (\mathbf{G}_1)_{\alpha\beta} \end{pmatrix}, \quad \mathbf{A}^{-1} = \begin{pmatrix} (\mathbf{S}_1)_{ij} & (\mathbf{\Delta}_{12})_{i\beta} \\ (\mathbf{\Delta}_{21})_{\alpha j} & (\mathbf{S}_1)_{\alpha\beta} \end{pmatrix} \tag{A2.1}$$

(here and in what follows the Latin and Greek indices run the values $1,..,m$ and $1,..,n$,

respectively.) Thus, to invert the matrix **A** means to find the blocks $\mathbf{S}_1$, $\mathbf{S}_2$, $\mathbf{\Delta}_{12}$ and $\mathbf{\Delta}_{21}$.

It is easy to check that the following inversion rules are valid [48].

1. If both square matrices $\mathbf{G}_1$ and $\mathbf{G}_2$ are not degenerate then

$$\mathbf{S}_1 = (\mathbf{G}_1 - \mathbf{\Sigma}_1)^{-1}, \quad \mathbf{S}_2 = (\mathbf{G}_2 - \mathbf{\Sigma}_2)^{-1},$$
$$\mathbf{\Delta}_{12} = \mathbf{G}_1^{-1}\mathbf{\Gamma}_{12}\mathbf{S}_2 = \mathbf{S}_1\mathbf{\Gamma}_{12}\mathbf{G}_2^{-1}, \quad \mathbf{\Delta}_{21} = \mathbf{G}_2^{-1}\mathbf{\Gamma}_{21}\mathbf{S}_1 = \mathbf{S}_2\mathbf{\Gamma}_{21}\mathbf{G}_1^{-1},$$
(A2.2)

where

$$\mathbf{\Sigma}_1 = \mathbf{\Gamma}_{12}\mathbf{G}_2^{-1}\mathbf{\Gamma}_{21}, \quad \mathbf{\Sigma}_2 = \mathbf{\Gamma}_{21}\mathbf{G}_1^{-1}\mathbf{\Gamma}_{12}.$$
(A2.3)

In particular, the matrices $\hat{\gamma}$ and $\hat{\gamma}_{BLS}$ have the form

$$A = \begin{pmatrix} a & Ta & -\dfrac{iq_\beta p}{q^2}a \\ Ta & T^2(a+b) & -\dfrac{iq_\beta pT}{q^2}a \\ -\dfrac{iq_\alpha p}{q^2}a & -\dfrac{iq_\alpha pT}{q^2}a & \dfrac{q_\alpha q_\beta}{q^2}\left(C - \dfrac{p^2}{q^2}\gamma_{pp}\right) + \left(\delta_{\alpha\beta} - \dfrac{q_\alpha q_\beta}{q^2}\right)W \end{pmatrix}$$
(A2.4)

The corresponding inverse matrix reads

$$A^{-1}(\mathbf{q}, p) = \begin{pmatrix} \dfrac{1}{a} + \dfrac{1}{b} - \dfrac{p^2}{q^2}\dfrac{1}{C} & -\dfrac{1}{bT} & \dfrac{iq_\beta p}{q^2 C} \\ -\dfrac{1}{bT} & \dfrac{1}{bT^2} & 0 \\ \dfrac{iq_\beta p}{q^2 C} & 0 & \dfrac{q_\alpha q_\beta}{q^2 C} + \dfrac{1}{W}\left(\delta_{\alpha\beta} - \dfrac{q_\alpha q_\beta}{q^2}\right) \end{pmatrix}.$$
(A2.5)

2. If one of the matrices (for definiteness, $\mathbf{G}_1$) is degenerate but the matrix $\mathbf{G}_1 - \mathbf{\Sigma}_1$ is not degenerate then

$$\mathbf{S}_1 = (\mathbf{G}_1 - \mathbf{\Sigma}_1)^{-1}, \quad \mathbf{S}_2 = \mathbf{G}_2^{-1} + \mathbf{G}_2^{-1}\mathbf{\Gamma}_{21}\mathbf{S}_1\mathbf{\Gamma}_{12}\mathbf{G}_2^{-1},$$
$$\mathbf{\Delta}_{12} = \mathbf{S}_1\mathbf{\Gamma}_{12}\mathbf{G}_2^{-1}, \quad \mathbf{\Delta}_{21} = \mathbf{G}_2^{-1}\mathbf{\Gamma}_{21}\mathbf{S}_1.$$
(A2.6)

Applying the formulas (A2.6) to inversion of the matrix $\hat{\mathbf{\Lambda}}_{BLS}$ defined by Eq. (5.11), we get

$$\hat{\mathbf{\Lambda}}_{BLS}^{-1} = \begin{pmatrix} \left(\dfrac{\Gamma_0(\nu+\mu)}{b^2} + \dfrac{c^2}{a}\right) & -\dfrac{c}{a} & \dfrac{in_\beta}{b} \\ -\dfrac{c}{a} & a^{-1} & 0 \\ \dfrac{in_\alpha}{b} & 0 & \dfrac{\delta_{\alpha\beta} - n_\alpha n_\beta}{\Gamma_0 \nu} \end{pmatrix}$$
(A2.7)

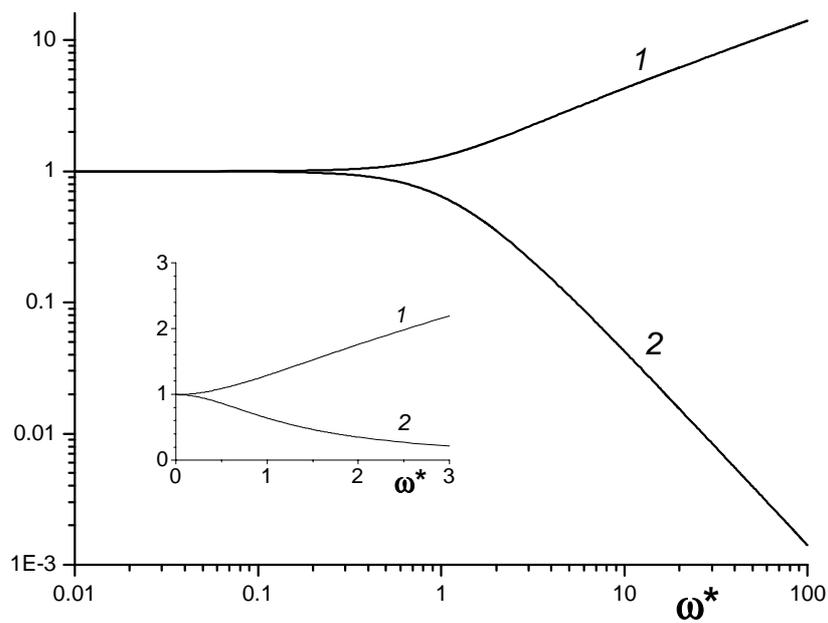

Fig 1. Frequency dependences of the normalized sound velocity $\tilde{u}$ (curve *1*) and sound absorption coefficient $\tilde{\Gamma}$ (curve *2*) on the normalized frequency $\omega*$ in double logarithmic co-ordinates. The initial parts of the curves are shown in inset in linear co-ordinates.